\documentclass[final,3p,times,twocolumn]{elsarticle}

\usepackage{amsmath, amsthm, amsfonts, amssymb,empheq, tensor,ulem,mathtools}
\usepackage[colorlinks=true,
            linkcolor=blue,
            filecolor=magenta,
            urlcolor=cyan,
            citecolor=magenta]{hyperref}

\graphicspath{{}}  
\DeclareMathAlphabet{\pazocal}{OMS}{zplm}{m}{n}
\newcommand{\La}{\mathcal{L}}

\DeclareMathAlphabet{\pazocal}{OMS}{zplm}{m}{n}

\newcommand{\Hb}{\mathcal{H}}

\usepackage{float}

\begin{document}

\hypersetup{
    colorlinks=true,
    linkcolor=blue,
    filecolor=magenta,      
    urlcolor=cyan,
    citecolor=magenta
}
\begin{frontmatter}
\title{Inflation in theories with broken diffeomorphisms}

\author[l1]{Antonio L. Maroto}
\ead{maroto@ucm.es}
\author[l1]{Prado Martín-Moruno}
\ead{pradomm@ucm.es}
\author[l1]{Miguel Orbaneja-Pérez\corref{c1}}
\ead{miorbane@ucm.es}
\cortext[c1]{Corresponding author}

\affiliation[l1]{organization={Departamento de Física Teórica and Instituto de Física de Partículas y del Cosmos (IPARCOS-UCM),\\ Universidad Complutense de Madrid},
            addressline={}, 
            city={Madrid},
            postcode={28040}, 
            state={},
            country={Spain}}

%% Abstract
\begin{abstract}
We analyze the impact of breaking diffeomorphism invariance in the inflaton sector. In particular, we consider inflaton models which are invariant under the subgroup of transverse diffeomorphisms and address the possibility of implementing a slow-roll phase. We obtain the corresponding expressions for relevant quantities such as the slow-roll parameters and the number of $e$-folds,  and derive the  primordial power-spectrum of curvature perturbations. The scalar spectral index features modifications which are confronted with CMB data from Planck and ACT. We study in detail the quadratic potential model, combining asymptotic and numerical analysis. We show that the post-inflationary behavior can be drastically different from the diffeomorphism-invariant case, exhibiting novel dynamical regimes.
\end{abstract}

%% Keywords

\end{frontmatter}

%% main text

\section{Introduction}\label{sec:INTRODUCTION TO TDIFF THEORY}

The standard $\Lambda\text{CDM}$ model has been astonishingly successful once modern cosmology became a precision science in the late twentieth century. The model has been shown to fit high precision data from early epochs of the universe since the time of light elements synthesis until today. Nevertheless, the incompleteness of $\Lambda\text{CDM}$ is widely recognized due to certain shortcomings which affect both early and late epochs of the evolution. At late times, establishing the nature of dark matter and dark energy is an open problem of cosmology. At early times, these loose ends can be mainly summarized \cite{Kolb:1990vq} as follows: the flatness problem, which relates to the very fine-tuned value of the spatial curvature; the horizon problem, related to the extreme isotropy of the temperature of the Last Scattering Surface; and the problem of the origin of the large-scale structure. These problems do not rule out $\Lambda\text{CDM}$, rather, they open the door to new components still compatible with our current understanding of the Universe.

The inflationary paradigm \cite{Guth:1980zm, Linde:1983gd, Riotto:2002yw} provides an elegant solution to the above mentioned issues at early times by introducing a short period of accelerated expansion in the very early universe. 
Inflation is typically implemented by a new sector containing an additional scalar field (``inflaton'') whose potential energy 
drives the acceleration during the so-called slow-roll phase. Although the 
inflaton sector is not necessarily contained within the Standard Model of elementary particles, models of inflation based on the Higgs field with suitable 
non-minimal couplings to gravity \cite{Bezrukov:2007ep} have been widely studied.
On the other hand, inflation can also be implemented by certain modifications of Einstein General Relativity in the high-curvature regime. This is the case of the well-known Starobinsky model \cite{Starobinsky:1980te}.  

The increasing precision of cosmological observations, mainly from CMB temperature and polarization, together with the most recent data of large-scale matter distribution from galaxy surveys, has allowed to set stringent constraints on different models of inflation. Thus, from Planck satellite \cite{Planck:2018jri} it has been possible to argue that the simplest inflaton models with renormalizable potential terms may be disfavored, with Starobinsky or Higgs-inflation models providing better fits to observations.  More recently, the results of the Atacama Cosmology Telescope (ACT) Data Release 6 (DR6) \cite{ACT:2025tim}, with better angular resolution, combined with DESI DR1 data 
suggest some tension with the previous Planck results within the $\Lambda$CDM scenario. In particular, the new data reduces the goodness of the fit of Starobinsky inflation. However, as noted in reference \cite{SPT-3G:2025bzu}, 
CMB and DESI data are known to be already in tension within 
$\Lambda$CDM, so that their use in a combined data analysis can be problematic. Anyway, possible extensions of the simplest inflaton models
including generic non-mininal couplings to gravity have been analyzed
as possible viable alternatives \cite{Kallosh:2025rni,Gao:2025onc,Kim:2025dyi}. 

In general, the inflationary models mentioned above are described by generally covariant
field theories, i.e. theories invariant under arbitrary diffeomorphisms (Diff). However, in recent years, the interest in gravity theories with broken diffeomorphisms has grown mainly motivated by the success of unimodular gravity \cite{Einstein:1919gv,Unruh.40.1048,Carballo-Rubio:2022ofy} as a possible solution to the so called vacuum-energy 
problem \cite{articleEllis}. In unimodular gravity, the metric determinant is considered as a fixed non-dynamical field with $g=1$ and, therefore, the Diff invariance is broken down to transverse diffeomorphisms (TDiff) and Weyl rescalings. Thus, unimodular gravity propagates the same number of 
degrees of freedom as General Relativity. On the other hand, TDiff gravity models beyond unimodular gravity have been studied in references \cite{Alvarez:2006uu,Pirogov:2011iq, Bello-Morales:2023btf,Bello-Morales:2024vqk}. Such theories propagate an additional scalar graviton mode and their  cosmological evolution was studied  in reference \cite{Bello-Morales:2023btf}. Furthermore, TDiff invariant theories with broken diffeomorphisms in the matter sector have been analyzed in references \cite{Maroto:2023toq,Jaramillo-Garrido:2023cor} for single scalar fields. Such theories behave as standard Diff models on small scales \cite{Maroto:2023toq}; however, on super-Hubble scales their behaviour can be drastically different, thus opening up a 
wide range of possibilities for cosmological model building. A particularly interesting case is 
the possibility of building  dark matter models with a simple free scalar fields \cite{Maroto:2023toq,Jaramillo-Garrido:2023cor}. In addition, the whole dark sector
of $\Lambda$CDM can also be described with a single scalar field with a 
canonical kinetic term  in this framework \cite{deCruzPerez:2025ytd}. An alternative unified TDiff model for the dark sector has been considered in reference \cite{Alonso-Lopez:2023hkx}. A general classification of 
single-field TDiff models based on their speed of sound  and equation of state was performed in \cite{Jaramillo-Garrido:2024tdv}. TDiff models with several scalar fields were analyzed in references \cite{Tessainer:2024ewm,Maroto:2025ife}, where the breaking of Diff symmetry naturally induces a coupling between the fields, thus providing a simple 
mechanism to generate interactions in the dark sector.
TDiff models for single abelian gauge fields were studied in \cite{Maroto:2024mkx} and their phenomenological implications for cosmic magnetic field 
evolution in reference \cite{Maroto:2024roe}. More recently, the equivalence between TDiff theories and certain classes of $k$-essence and mimetic models
has been established \cite{BeltranJimenez:2025puw}.

In this work we will examine the consequences of breaking Diff symmetry down to TDiff in the inflaton sector.  In particular, the fundamental ideas of inflation will be summarized in section \ref{sec:INFLATION}
and then we will lay the groundwork of TDiff theories in \ref{sec:TDIFF FRAMEWORK}. In section \ref{sec: SLOW-ROLL IN TDIFF THEORIES}, we will apply the TDiff framework to inflation, revisiting definitions and involved quantities. Section \ref{sec:PowerSpectrum} is focused on obtaining the relevant observables of our model.
In section \ref{sec:Metric pert}, we will study the metric perturbations in the TDiff theory. In section \ref{sec:QUANTIZATION}, we will derive the primordial power-spectrum of curvature perturbations.
Then, in section \ref{sec:phenomenology}, we will compare the predictions of the TDiff slow-roll inflationary models with the available observational data.
Focusing on the case of power law potentials, we will find the spectral index and the tensor-to-scalar ratio in section \ref{sec:power-law} and we will compare the results with the experimental data in section \ref{sec:Results comparison}. In section \ref{sec:TDIFF QUADRATIC POTENTIAL}, we analyze the dynamical system for our model discussing also the post-inflationary phase. We comment on the general characteristic of the strong TDiff regime in section \ref{sec:Strong TDiff regime}, discussing the corresponding phase portraits \ref{sec:portraits}. We also present the detailed analysis of a particular model in section \ref{sec:Example of TDiff dynamical system}. We present the conclusions in section \ref{sec:Conclusions}. Finally, we discuss in deeper detail the propagating degrees of freedom and quadratic action for perturbations in \ref{sec:ApA}.

Throughout this manuscript we will use the metric signature $(+,-,-,-)$ and natural units $\hbar = c = 1$.
\section{PRELIMINARY CONCEPTS}\label{sec:PRELIMINARIES}
In this section, we present a summary of the theoretical foundations upon which the TDiff inflationary model addressed in this paper is constructed. First, a brief overview of the inflationary paradigm is provided. Then, we include a summary of TDiff scalar field theories, emphasizing their main characteristics.

\subsection{Standard cosmic inflation}\label{sec:INFLATION}
One of the simplest types of inflationary models is based on the existence of a real scalar field known as ``inflaton", denoted by $\phi$, which is minimally coupled to gravity via the following action \cite{Baumann:2022mni, DiMarco:2024yzn}
\begin{equation}
    S^{\text{Diff}}_\phi = \int d^4x \sqrt{g} \left[\frac{1}{2}g^{\mu\nu}\partial_\mu\phi\partial_\nu\phi-V(\phi)\right]
    \label{eq:SDiff},
\end{equation}
where $g=|\text{det}(g_{\mu\nu})|>0$ is the absolute value of the metric determinant and $V(\phi)$ the potential function. The potential energy density of the inflaton field will be responsible for the accelerated expansion. In addition, a standard inflationary evolution for the scale factor involves a quasi-de Sitter dependence, $a(t)\simeq e^{H_It}$, with a nearly constant Hubble parameter, $H_I$. Taking into account a Friedmann-Lemaître-Robertson-Walker (FLRW) background for a homogeneous, isotropic and spatially flat universe, given by 
\begin{equation}
    ds^2=dt^2-a(t)^2\left(\,d\vec{x}\,\right)^2
    \label{eq:ds2 FRLW},
\end{equation}
the equation of motion (EoM) of the field then reads
\begin{equation}
    \ddot\phi+3H\dot\phi+V'(\phi)=0,
    \label{eq:Diff eom}
\end{equation}
where $\dot{}=d/dt$, the prime denotes derivative with respect to the argument and $H=\dot{a}/a$.
The resulting Friedmann and conservation equations read
\begin{align}
        H^2 & = \frac{8\pi G}{3}\rho \label{eq:FLRW} ,\\
        \dot{\rho} & =-3H\left(\rho+p\right) \label{eq:cons},
\end{align}
where $\rho$ and $p$ are the energy density and the pressure of $\phi$, respectively, given by
\begin{subequations}
    \begin{align}
        \rho & =\frac{1}{2}\dot{\phi^2}+V(\phi), \label{eq:rho Diff}\\
     p & =\frac{1}{2}\dot{\phi^2}-V(\phi) . \label{eq:p Diff}
\end{align}
\end{subequations}
The combination of equations \eqref{eq:FLRW} and \eqref{eq:cons} yields
\begin{equation}
   \dot{H} = -4\pi G \left(\rho + p\right).
  \label{eq:dotH}
\end{equation}
As a practical matter, a nearly constant $H$ ($\dot{H}\simeq0$) could be achieved if $p\simeq-\rho$, as seen above, which translates into the standard slow-roll condition $\dot{\phi}^2\ll V(\phi)$, given \eqref{eq:rho Diff} and \eqref{eq:p Diff}. This condition can be obtained if the friction term in equation \eqref{eq:Diff eom} dominates, as for an overdamped oscillator, that is $\ddot\phi\ll \{V', 3H\dot\phi\}$. This is the so-called slow-roll regime and, throughout the rest of the work, we will be using the symbol $\simeq$ to express the application of this regime.

In order to obtain the slow-roll conditions, the following slow-roll parameters are defined:
\begin{align}
        \varepsilon & = -\frac{\dot H}{H^2}
       \simeq\frac{1}{16\pi G}\left[\frac{V^\prime(\phi)}{V(\phi)}\right]^2, \label{eq:Diff epsilon}
        \\
        \eta & = \varepsilon+\delta 
        =\varepsilon  -\frac{\ddot\phi}{\dot\phi H}\simeq
        \frac{1}{8\pi G} \frac{V^{\prime\prime}(\phi)}{V(\phi)}. \label{eq:Diff eta}
\end{align}
Thus, the slow-roll conditions are $\varepsilon\ll1$ and $|\eta|\ll1$, which establishes constraints on $V(\phi)$. On the one hand, the condition on $\varepsilon$ indicates that the expansion rate of the universe should be nearly constant during inflation ($\dot H\simeq0$); on the other hand, the condition on $\eta$ implies that the friction term governs the field EoM \eqref{eq:Diff eom}. This slow-roll regime is no longer valid once $\text{max}\{\varepsilon(\phi_f),|\eta(\phi_f)|\}=1$, with this condition signaling the end of inflation for $\phi_f$.

Lastly, to measure the duration of the inflationary epoch, the number of $e$-folds  is commonly defined as $N=\ln\frac{a_f}{a_i}$, where $a_{i,f}$ are the scale factor at the beginning and end of this epoch. This quantity can be expressed as
\begin{equation}
         N=%\int^{a_f}_{a_i}\frac{da}{a} 
         \int^{\phi_f}_{\phi_i}d\phi
         \frac{H(\phi)}{\dot\phi}
        \simeq-8\pi G\int^{\phi_f}_{\phi_i}d\phi\frac{V(\phi)}{V^{\prime}(\phi)}
    \label{eq:Diff efolds},
\end{equation}
and, typically, $N\gtrsim50$ \cite{Liddle:1994dx,DiMarco:2024yzn} is required in order for inflation to solve the horizon and flatness problems. After inflation, a reheating phase should occur in which the inflaton energy is transferred to a hot thermal plasma in which nucleosynthesis can take place.
\subsection{TDiff scalar field theory}\label{sec:TDIFF FRAMEWORK}
TDiff transformations \cite{{Maroto:2023toq}} are coordinate transformations which keep both the metric determinant $g$ and volume element $d^4x$ invariant. Under a general infinitesimal transformation $\hat{x}^\mu=x^\mu+\xi^\mu(x)$, these quantities transform as
\begin{align}
    \hat{g}(\hat{x}) & = \left[1-2\partial_\mu  \xi^\mu(x)+\mathcal{O}(\xi^2)\right]g(x) = g(x),\label{eq:TDiff g invariance}\\
     d^4\hat{x} & = \left[1+\partial_\mu  \xi^\mu(x)+\mathcal{O}(\xi^2)\right]d^4x = d^4x ;\label{eq:TDiff d4x invariance}
\end{align}
so TDiff transformations have to satisfy the TDiff condition:
\begin{equation}
    \partial_\mu  \xi^\mu(x) = 0.
    \label{eq:TDiff cond}
\end{equation}
Thus, to lowest order in metric derivatives, any type of action of the form
\begin{equation}
    S^{\text{TDiff}}_\psi[\psi,g_{\mu\nu}] = \int d^4x f(g) \La \left(\psi,\partial_\mu\psi,g_{\mu\nu}\right)
    \label{eq:STDiff},
\end{equation}
with $\La$ a Diff invariant Lagrangian, will be invariant under TDiff transformations. Here, $f(g)$ is an arbitrary function of the metric determinant and $\psi$ is a generic field. From now on, we shall identify the field $\psi$ as the inflaton $\phi$. Assuming a universal volume function for the kinetic and potential terms and also a canonical kinetic term for the field, we can write the inflaton action as
\begin{equation}
    S_\phi = \int d^4x f(g)\left[\frac{1}{2}g^{\mu\nu}\partial_\mu\phi\,\partial_\nu\phi-V(\phi)\right],
    \label{eq:actionphi}
\end{equation}
where we shall assume a positive function $f(g)$ to avoid instabilities \cite{Maroto:2023toq,Jaramillo-Garrido:2023cor}.

An alternative formulation of this TDiff theory can be implemented through the so-called covariantization procedure in which an additional Stueckelberg-like field $A^\mu$ is introduced to restore the full Diff invariance. The corresponding covariantized action \cite{Jaramillo-Garrido:2024tdv} can accordingly be written as 
\begin{equation}
    S_{\phi}^{\text{cov}} = \int d^4x \sqrt{g}H_K(Y)\left[\frac{1}{2}g^{\mu\nu}\partial_\mu\phi\partial_\nu\phi-V(\phi)\right]
    \label{eq:SphiDiff},
\end{equation}
where we have defined $Y=\nabla_\mu A^\mu$ and the function
\begin{equation}
    H_K(Y)= Yf(Y^{-2}),
    \label{eq:Hk}
\end{equation}
so that the previous action \eqref{eq:actionphi} is recovered in the TDiff coordinate frame in which
\begin{equation}
    Y=\frac{1}{\sqrt{g}}.
    \label{eq:Y g}
\end{equation}
We stress that, in this notation, the Diff case $f(g)=\sqrt{g}$ is readily recovered via the function $H_K(Y)=1$.

On the one hand, variations of the action \eqref{eq:SphiDiff} with respect to the scalar field yield the EoM
\begin{equation}                    
\nabla_\mu\left[H_K(Y)g^{\mu\nu}\nabla_\nu\phi\right]+H_K(Y)V^\prime(\phi)=0,
 \label{eq:EoM phiDiff}
\end{equation}
where, once again, the prime denotes derivative with respect to its argument. On the other hand, variations with respect to the vector field $A^\mu$ lead to
\begin{equation}
    \partial_\alpha\left[H_K^\prime(Y)\left(X-V\right)\right]=0,
    \label{eq:Tcov deriv}
\end{equation}
with the corresponding notation
\begin{equation}
    X=\frac{1}{2}g^{\mu\nu}\partial_\mu\phi\partial_\nu\phi.
    \label{eq:X cov}
\end{equation}
Equivalently, the equation \eqref{eq:Tcov deriv} can be recast as
\begin{equation}
    H_K'(Y)(X-V)=-\frac{c_\rho}{2},
    \label{eq:Tcov const}
\end{equation}
being $c_\rho$ a constant. We shall refer to this last equation hereinafter as the TDiff constraint, as it fixes the new physical degree of freedom which arises due to the symmetry breaking.

Lastly, let us consider the total action $S=S_{EH}+S^{\text{cov}}_\phi$, with $S_{EH}$ the usual Einstein-Hilbert action for General Relativity. We remark that we are breaking down the Diff invariance through the matter sector, thus, the resulting Einstein equations take the common form
\begin{equation}
    R_{\mu\nu}-\frac{1}{2}g_{\mu\nu}R=8\pi GT_{\mu\nu},
\end{equation}
where the energy-momentum tensor (EMT) is defined in the usual way as
\begin{equation}
    T^{\mu\nu}=-\frac{2}{\sqrt{g}}\frac{\delta  S_{\phi}^{\text{cov}}}{\delta g_{\mu\nu}}.
\end{equation}
Thus, the EMT reads
\begin{equation}
\begin{split}
     T_{\mu\nu}& =H_K(Y)\partial_\mu\phi\partial_\nu\phi\, \\&\,\,\,\,\,\,\,-\left[H_K(Y)-YH_K'(Y)\right](X-V)g_{\mu\nu}.
\end{split}
\end{equation}
Substituting now equation \eqref{eq:Tcov const} into the last one, we are then able to write
\begin{equation}
\begin{split}
    T_{\mu\nu}&=H_K(Y)\partial_\mu\phi\partial_\nu\phi\,\\&\,\,\,\,\,\,\,-\left[H_K(Y)(X-V)+Y\frac{c_\rho}{2}\right]g_{\mu\nu}.
\end{split}
\end{equation}
We can easily recognize the EMT of a perfect fluid $T_{\mu\nu}=(\rho+p)u_\mu u_\nu-pg_{\mu\nu}$ if we properly consider the fluid velocity $u_\mu=\frac{\partial_\mu\phi}{\sqrt{2X}}$ \cite{Jaramillo-Garrido:2024tdv}. Thus, the energy density and pressure are just 
\begin{subequations}
    \begin{align}
        \rho & =H_K(Y)(X+V)-\frac{c_\rho}{2}Y, \label{eq:rho TDiff cov}\\
     p & =H_K(Y)(X-V)+\frac{c_\rho}{2}Y . \label{eq:p TDiff cov}
\end{align}
\end{subequations}
We remark the following relation for later convenience throughout the rest of the work:
\begin{equation}
    \rho+p=2XH_K(Y).
    \label{eq:sum rhop}
\end{equation}

From now on, we focus on a homogeneous field, $\phi=\phi(t)$, in a cosmological background described by the FLRW metric \eqref{eq:ds2 FRLW}. In this background the EoM \eqref{eq:EoM phiDiff} for the inflaton field is
\begin{equation}
    \ddot{\phi} + \left[3H+\frac{H'_K(Y)}{H_K(Y)}\dot Y\right]\dot{\phi} + V^{\prime}\left(\phi\right) = 0
    \label{eq:EoM TDiff0}.
\end{equation}
In addition, the constraint equation \eqref{eq:Tcov const} reads
\begin{equation}
     H_K'(Y)\left(\frac{1}{2}\dot\phi^2-V\right)=-\frac{c_\rho}{2}.\label{const}
\end{equation}
Einstein equations still yield the Friedmann equation \eqref{eq:FLRW} and the previously seen relation \eqref{eq:dotH}. That expression can be combined with the relation \eqref{eq:sum rhop} to find
\begin{equation}
    \dot{H} = -4\pi G H_K(Y)\,\dot\phi^2.
    \label{eq:Hdot cov}
\end{equation}
In the next section we will start to examine the main changes that arise in inflation in the  TDiff framework.
\section{SLOW-ROLL TDiff INFLATION}\label{sec: SLOW-ROLL IN TDIFF THEORIES}
Let us now consider that the cosmic inflationary phase is driven by a TDiff scalar field. Moreover, for the sake of simplicity, we will take a power-law for the TDiff volume function, that is,
\begin{equation}
    f(g)=g^\alpha
    \label{eq:plaw}.
\end{equation}
Therefore, making use of \eqref{eq:Y g}, the function \eqref{eq:Hk} becomes
\begin{equation}
    H_K(Y)=Y^{1-2\alpha}.
    \label{eq:Hk plaw}
\end{equation}
It can be noted that the Diff case is recovered by making $\alpha=1/2$. Taking into account equation \eqref{eq:Hk plaw}, the energy density \eqref{eq:rho TDiff cov} and pressure \eqref{eq:p TDiff cov} take the form
\begin{subequations}
    \begin{align}
        \rho & =Y^{1-2\alpha}\left[(1-\alpha)\dot\phi^2+2\alpha V(\phi)\right], \label{eq:rho TDiff plaw}\\
     p & =Y^{1-2\alpha}\left[\alpha\dot\phi^2-2\alpha V(\phi)\right] , \label{eq:p TDiff plaw}
\end{align}
\end{subequations}
so that the equation of state (EoS) parameter reads
\begin{equation}
    w_\phi=-1+\frac{\dot\phi^2}{(1-\alpha)\dot\phi^2+2\alpha V}.\label{eq:w}
\end{equation}

Let us also consider that the inflaton field evolves within the slow-roll regime. In order to have positive energy density during slow-roll, the condition $\alpha>0$ must be required. Proceeding now as in section \ref{sec:INFLATION}, slow-roll parameters shall be firstly computed. Making use of equations \eqref{eq:FLRW}, \eqref{eq:rho TDiff plaw} and \eqref{eq:Hdot cov}, the first slow-roll parameter \eqref{eq:Diff epsilon} is just
\begin{equation}
    \varepsilon = \frac{3}{2}\frac{\dot{\phi}^2}{(1-\alpha)\,\dot\phi^2+2\alpha V(\phi)}\simeq\frac{3}{4\alpha}\frac{\dot{\phi}^2}{V(\phi)}
    \label{eq:eps slr cov},
\end{equation}
where the last expression is valid during the slow-roll regime. This parameter can be used to recast the EoS parameter \eqref{eq:w} as 
\begin{equation}
    w_\phi\simeq-1+\frac{2}{3}\varepsilon.
    \label{eq:EoS phi}
\end{equation}
We can also find the time dependence of the scale factor. For later convenience, we define the conformal time $\tau$ as
\begin{equation}
    dt=a(\tau)d\tau,
    \label{eq:conformal time}
\end{equation}
so that the definition of the $\varepsilon$ parameter \eqref{eq:Diff epsilon} becomes
\begin{equation}
    \varepsilon=1-\frac{\Hb'}{\Hb^2},
    \label{eq:eps conformal time}
\end{equation}
where we denote $'=d/d\tau$, unless otherwise stated, and the conformal Hubble parameter is defined as $\Hb=a'/a=aH$. Taking into account that during slow-roll $\varepsilon'\sim\mathcal{O}(\varepsilon^2)$ \cite{Liddle:1994dx,Baumann:2022mni}, we can assume that $\varepsilon$ varies slowly and integrate the above relation, so that
\begin{equation}
  \Hb\simeq\frac{1}{(1-\varepsilon)(-\tau)}. 
    \label{eq:Hetaeps}
\end{equation}
This also yields $a\propto|\tau|^{-(1+\varepsilon)}$ after integrating and expanding for $\varepsilon\ll1$. We remark that the conformal time $\tau$ is negative during slow-roll.

In order to express $\varepsilon$ in terms of the potential and its derivatives, we substitute the power-law function \eqref{eq:Hk plaw} into the EoM \eqref{eq:EoM TDiff0}, so that
\begin{equation}
    \ddot{\phi} + \left[3H+(1-2\alpha)\frac{\dot Y}{Y}\right]\dot\phi+V^{\prime}\left(\phi\right) = 0.
    \label{eq:EoM TDiff}
\end{equation}
In the TDiff overdamped regime, that is, for $\ddot{\phi}\ll\left \{3H\dot{\phi}, \frac{\dot Y}{Y}\dot\phi,V^{\prime}(\phi) \right\}$, one has
\begin{equation}
    V^{\prime}(\phi)\simeq-\left[3H+(1-2\alpha)\frac{\dot Y}{Y}\right]\dot\phi
    \label{eq:EoM slr}.
\end{equation}
Additionally, the constraint \eqref{eq:Tcov deriv} can also be recast to the leading order in the slow-roll approximation as follows
\begin{equation}
    2\alpha\frac{\dot Y}{Y}V(\phi)\simeq V'(\phi)\dot\phi.
    \label{eq: dep YV}
\end{equation}
Combining this last equation with expressions \eqref{eq:EoM slr} and \eqref{eq:eps slr cov}, we obtain the evolution of $Y$
\begin{equation}
    \frac{\dot{Y}}{Y} \simeq -2\varepsilon H
    \label{eq:dep Y-a},
\end{equation}
which readily implies, since the $\varepsilon$ parameter is constant to first order in the slow-roll approximation, $Y\propto a^{-2\varepsilon}$. Essentially, this relation features a slow evolution for the field $Y$ during slow-roll. This argument also stands for the function \eqref{eq:Hk plaw}, $H_K(Y)\propto a^{(4\alpha-2)\varepsilon}$.
Finally, we can now substitute equation \eqref{eq:dep Y-a} back into the EoM \eqref{eq:EoM slr} and find
\begin{equation}
    V'(\phi)\simeq-3H\dot\phi,
    \label{eq:EoM simp}
\end{equation}
to leading order in $\varepsilon$.
Now, taking into account equations \eqref{eq:FLRW}, \eqref{eq:rho TDiff plaw} and \eqref{eq:EoM simp}, the slow-roll parameter \eqref{eq:eps slr cov} can be recast as
\begin{equation}
    \varepsilon \simeq \frac{1}{64\pi G\alpha^2}Y^{2\alpha-1} \left(\frac{V^{\prime}}{V}\right)^2
    \label{eq:prim par slr}.
\end{equation}
For convenience, it is also worthwhile to relate the field $Y$ with $V(\phi)$. To this end, during slow roll we can express equation \eqref{eq: dep YV} as
\begin{equation}
   \frac{\dot Y}{Y}\simeq \frac{1}{ 2\alpha}\frac{\dot V}{V}
   \quad\Rightarrow\quad Y\propto V^{\frac{1}{2\alpha}}\label{eq: dep YV chain}
\end{equation}

On the other hand, we can differentiate equation \eqref{eq:EoM simp} to substitute it into 
the definition \eqref{eq:Diff eta}. Using again \eqref{eq:rho TDiff plaw} and \eqref{eq:EoM simp} one finds
\begin{equation}
    \eta\simeq \frac{V''}{3H^2}
    \simeq \frac{1}{16\pi G\alpha}Y^{2\alpha-1}\left(\frac{V^{\prime\prime}}{V}\right).
    \label{eq:sec par slr}
\end{equation}
In light of the above results, we stress that the expressions for first and second slow-roll parameters remarkably present new TDiff pre-factors, i. e., $H_K^{-1}(Y)=Y^{2\alpha-1}$, in contrast to the well-known Diff case \eqref{eq:Diff epsilon} and \eqref{eq:Diff eta}.

Finally, the end of the slow-roll regime is fixed by the condition $\text{max}\{\varepsilon(\phi_f),|\eta(\phi_f)|\}=1$. Analogously to the Diff case, one can obtain
the number of $e$-folds using the simplified EoM \eqref{eq:EoM simp} in the definition of $N$ \eqref{eq:Diff efolds}. Then, taking into account equations \eqref{eq:FLRW} and \eqref{eq:rho TDiff plaw}, and substituting the dependence of $Y$ obtained from equation \eqref{eq: dep YV chain}, one finds
\begin{equation}
    N\simeq-16\pi G\alpha Y_f^{1-2\alpha}V_f^{1-\frac{1}{2\alpha}} \int^{\phi_f}_{\phi_i}d\phi\frac{V^{\frac{1}{2\alpha}}(\phi)}{V^{\prime}(\phi)}.
   \label{eq:TDiff efolds}
\end{equation}
where $V_f= V(\phi_f)$ and $Y_f= Y(t_f)$. The comparison with the Diff expression \eqref{eq:Diff efolds} showcases again new pre-factors that could modify the final value for the number of \textit{e}-folds.
\section{PRIMORDIAL POWER-SPECTRUM}\label{sec:PowerSpectrum}
Let us now explore how the TDiff theory affects the metric perturbations. The quantization of these perturbations will allow us in the end to compute the primordial power-spectrum.
\subsection{Metric perturbations}\label{sec:Metric pert}
The most general form of the flat FLRW metric with scalar perturbations in conformal time \cite{Mukhanov:2005sc} has the following line element:
\begin{equation}
   \begin{split}
    ds^2 = a^2(\tau)\left\{\left(1+2\Phi\right)d\tau^2-2\partial_iBd\tau dx^i-\right.\\
    \left.\left[(1-2\Psi)\delta_{ij}+\partial_i\partial_j E\right]dx^idx^j\right\}
    \end{split}
\end{equation}
The perturbed scalar field is given by
\begin{equation}
    \phi=\phi_0(\tau)+\delta\phi(\tau,\vec{x}).
\end{equation}
We can also write the contributions for each component of the action \eqref{eq:SphiDiff} to first order in perturbations, namely,
\begin{align}
    X & =X_0+\delta X, \label{eq:X0 dX}\\
    V &= V_0+\delta V,\label{eq:V0 dV}\\
    Y & =Y_0+\delta \label{eq:Y0 dY}Y;
\end{align}    
where the subindex $0$ means evaluation at the background value.
Taking into account the above expressions, the EoM \eqref{eq:EoM TDiff} for the background field $\phi_0$ in conformal time \eqref{eq:conformal time} is just
\begin{equation}
    \phi_0'' + \left[2\Hb+(1-2\alpha)\frac{Y'_0}{Y_0}\right]\phi_0'+a^2V^{\prime}\left(\phi_0\right) = 0.
    \label{eq:EoM TDiff backg}
\end{equation}
where we denote $V'(\phi_0)=\left.dV(\phi)/d\phi\right\rvert_{\phi_0}$. The resulting Friedmann and conservation equations for the background in conformal time \eqref{eq:conformal time} are easily obtained from equations \eqref{eq:FLRW} and \eqref{eq:cons}:
\begin{align}
        \Hb^2 & = \frac{8\pi G}{3}a^2\rho \label{eq:FLRW conf} ,\\
        \rho' & =-3\Hb\left(\rho+p\right) \label{eq:cons conf}.
\end{align}   
When combined, they also yield
\begin{equation}
    \Hb'-\Hb^2=-4\pi G a^2(\rho+p).
    \label{eq:FLRW and cons conf}
\end{equation}
On the one hand, we can explicitly derive the expressions \eqref{eq:X0 dX}--\eqref{eq:Y0 dY} to first order in perturbations. Firstly, the kinetic term \eqref{eq:X cov} reads
\begin{align}
    X_0 &= \frac{1}{2a^2}\phi_0'^2, 
    \label{eq:X0}
    \\
    \delta X &= -\frac{\Phi}{a^2}\phi_0'^2+\frac{1}{a^2}\phi_0'\delta\phi'=2X_0\left(\frac{\delta\phi'}{\phi_0'}-\Phi\right).
    \label{eq:deltaX}
\end{align}
Secondly, for the potential term we obtain
\begin{align}
   % V_0 &= V(\phi_0), 
    %\label{eq:V0}
    %\\
    \delta V &= V'(\phi_0)\delta\phi.
    \label{eq:deltaV}
\end{align}
Finally, from the constraint equation \eqref{eq:Tcov const} we obtain for the field $Y$
\begin{equation}
    \frac{\delta Y}{Y_0} = \frac{1}{2\alpha}\frac{\delta X-\delta V}{X_0-V_0}.
    \label{eq:deltaY}
\end{equation}

On the other hand, the perturbed Einstein equations $8\pi  G \delta\tensor{T}{^\mu_\nu}=\delta \tensor{G}{^\mu _\nu}$ read in components
\begin{subequations}
    \begin{align}
        \begin{split}
        4\pi Ga^2\delta\tensor{T}{^0_0} & =\Hb\left[\nabla^2(E'-B)-3\Psi'\right]\\&\,\,\,\,\,\,\,+\nabla^2\Psi-3\mathcal{H}^2\Phi ; \label{eq:G00}
        \end{split}
        \\
    4\pi Ga^2\delta\tensor{T}{^0_i} & = \partial_i\left(\Psi'+\Hb\Phi\right) ; \label{eq:G0i}\\
    \begin{split}
        -4\pi Ga^2\delta\tensor{T}{^i_j} & =\left[\Psi''+\Hb\left(\Phi'+2\Psi'\right)\right.\\&\,\,\,\,\,\,\,+\left(2\Hb'+\Hb^2\right)\Phi-\left.\nabla^2D\right]\tensor{\delta}{^i_j}\\&\,\,\,\,\,\,\,+\partial_i\partial_jD ; \label{eq:Gij}
    \end{split}
\end{align}
\end{subequations}
where we have defined
\begin{equation}
    D=\frac{1}{2}(\Psi-\Phi)+\frac{1}{2}(E''-B')+\mathcal{H}(E'-B)\label{eq:Ddef}
\end{equation}
and where the indices are being raised and lowered with the metric of the spatial sections $\tensor{\delta}{^i_j}$.
Note that the perturbed geometrical sector, i.e. the Einstein Tensor $\delta\tensor{G}{^\mu_\nu}$, is identical to the one in General Relativity. This is due to the fact that the symmetry breaking occurs only through the matter sector, that is, the EMT $\delta\tensor{T}{^\mu_\nu}$.

\begin{figure}[t]
    \centering
    \includegraphics[width=\linewidth]{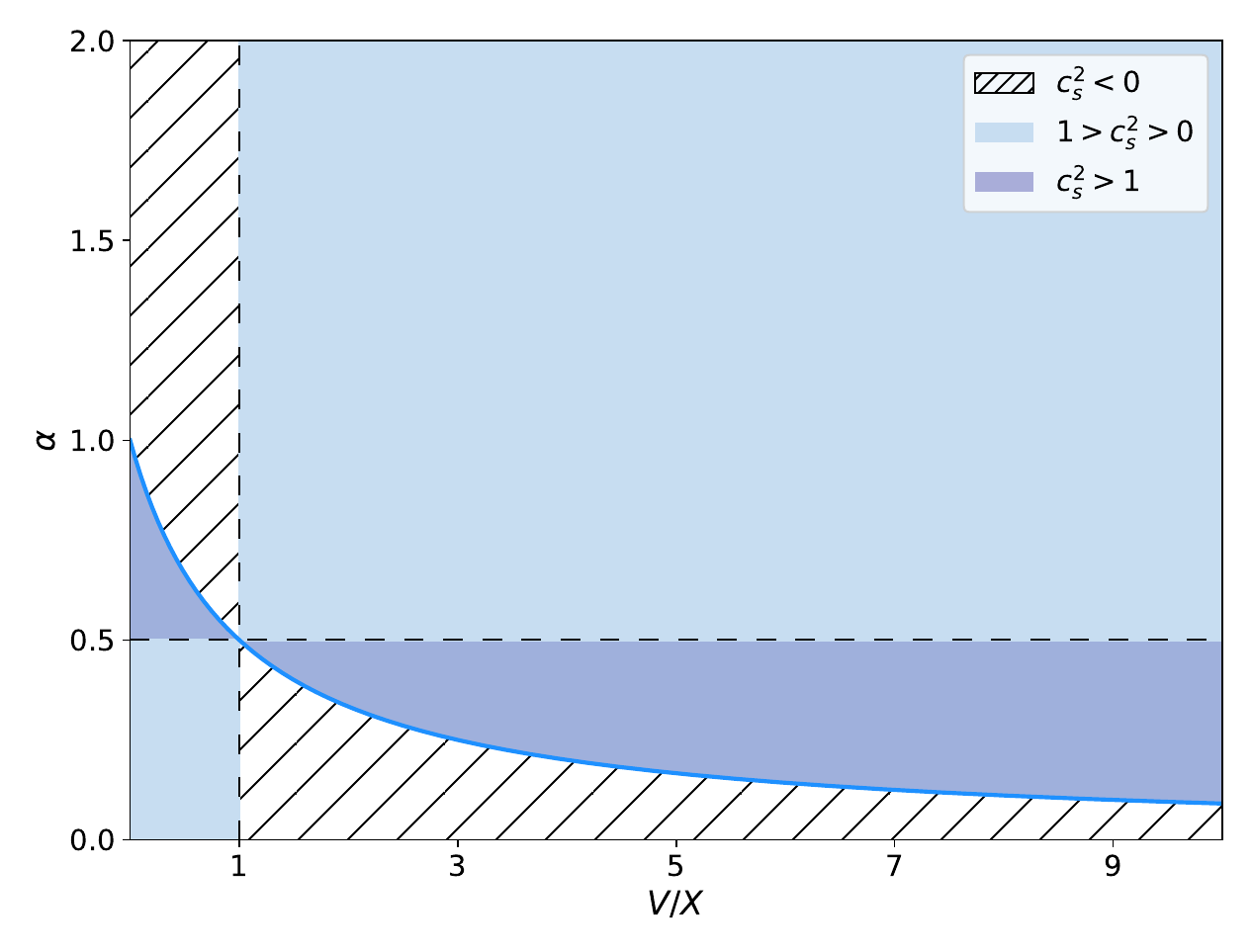}
    \caption{
    Excluded and allowed values of $\alpha$ as a function of  $V/X$. The striped area represents the excluded values, which corresponds to $c_s^2<0$. The light blue colored region represents  values with $1>c_s^2>0$ and the dark blue colored area corresponds to values with $c_s^2>1$. Finally, the blue colored line is the limiting region where $c_s^2$ in \eqref{eq:cs2} changes sign. Notice that on the dashed line with $\alpha=1/2$ we obtain the Diff value $c_s^2=1$, whereas on the vertical line with $V/X=1$, $c_s^2=0$. Notice also that potential dominated models with $V/X>1$ and $\alpha>1/2$ are in the $1>c_s^2>0$ region.}
    \label{fig:cs2}
\end{figure}

After imposing the perturbed constraint \eqref{eq:deltaY}, the components of the EMT perturbation read
\begin{subequations}
    \begin{align}
    \begin{split}
    \delta\tensor{T}{^0_0} & =\delta\rho=Y_0^{1-2\alpha}\Biggr[\delta X+\delta V\\
    &\,\,\,\,\,\,\,\,\,\,\,\,\,\,\,\,\,\,\,\,\,+\frac{1-2\alpha}{\alpha\left(1-\frac{V_0}{X_0}\right)}(\delta X-\delta V)\Biggr] ; \label{eq:deltaT00}
    \end{split}
    \\
    \delta\tensor{T}{^0_i} & =Y_0^{1-2\alpha}\frac{\phi_0'}{a^2}\partial_i\delta\phi ; \label{eq:deltaT0i}\\
    \delta\tensor{T}{^i_j} & =-Y_0^{1-2\alpha}(\delta X-\delta V)\tensor{\delta}{^i_j} ; \label{eq:deltaTij}
\end{align}
\end{subequations}
For later convenience, we can rewrite some of the previous components in an alternative way. In the case of the energy density, we can firstly define the effective speed of sound by differentiating the energy density \eqref{eq:rho TDiff cov} and the pressure \eqref{eq:p TDiff cov} at constant $\phi$, where we are making use of the constraint \eqref{eq:Tcov const} in order to consider the function $X=X(Y,\phi)$ \cite{Jaramillo-Garrido:2023cor}. Thus, we find
\begin{equation}
    c_s^2=\left.\frac{p_Y}{\rho_Y}\right\rvert_\phi=\frac{1}{1+\frac{1-2\alpha}{\alpha\left(1-\frac{V}{X}\right)}},
    \label{eq:cs2}
\end{equation}
where the subindex $Y$ stands for the partial derivative $\partial/\partial Y$. Notice that, in general, it is not always guaranteed that $c_s^2$ is positive for all values of $(V/X,\alpha)$. In Figure \ref{fig:cs2}, we show  the different regions in the $(V/X,\alpha)$ plane.  Regions with $c_s^2<0$ would lead to gradient instabilities in the scalar sector and should be excluded.

In addition, the conservation equation \eqref{eq:cons conf} for the background can be recast as
\begin{equation}
    -3\Hb\left(\rho_0+p_0\right)=\rho_0'=\frac{\partial\rho_0}{\partial X_0}X_0'+\frac{\partial\rho_0}{\partial Y_0}Y_0'+\frac{\partial\rho_0}{\partial \phi_0}\phi_0' \label{eq:cons conf backg}.
\end{equation}
where we have the derivatives
\begin{align}
    \frac{\partial\rho_0}{\partial X_0} & = H_K(Y_0),
    \label{eq:rhoX0}
    \\
    \frac{\partial\rho_0}{\partial Y_0} & = 2X_0H_K'(Y_0),
    \label{eq:rhoY0}
    \\
    \frac{\partial\rho_0}{\partial \phi_0} & = H_K(Y_0)V'(\phi_0).
    \label{eq:rhophi0}
\end{align}
Making use of the constraint \eqref{eq:Tcov const}, we can write
\begin{equation}
\begin{split}
        Y_0' & =\frac{\partial Y_0}{\partial X_0}X_0'+\frac{\partial Y_0}{\partial V_0}\frac{\partial V_0}{\partial\phi_0}\phi_0'=\\
        & = \frac{Y_0}{2\alpha(X_0-V_0)}\left[X_0'-\frac{\partial\rho_0}{\partial \phi_0}\frac{\phi_0'}{H_K(Y_0)}\right],
\end{split}
\label{eq:Y0'}
\end{equation}
where we have considered the relation \eqref{eq:rhophi0}. These steps allow us to perform the following rearrangements, taking into account \eqref{eq:cons conf backg}, so that
\begin{equation}
     \frac{\partial\rho_0}{\partial \phi_0} = -\frac{3\Hb}{\phi_0'}\left(\rho_0+p_0\right)-\frac{\partial\rho_0}{\partial X_0}\frac{X_0'}{\phi_0'}-\frac{\partial\rho_0}{\partial Y_0}\frac{Y_0'}{\phi_0'}.
     \label{eq:rhophi0 cons v1}
\end{equation}
We can now use the equation \eqref{eq:Y0'} and bear in mind the effective speed of sound \eqref{eq:cs2}, in order to obtain:
\begin{equation}
    \frac{\partial\rho_0}{\partial \phi_0}= -\frac{3\Hb}{\phi_0'}\left(\rho_0+p_0\right)\frac{c_s^2}{2c_s^2-1}-\frac{H_K(Y_0)}{\phi_0'}\frac{1}{2c_s^2-1}X_0'.
    \label{eq:rhophi0 cons v2}
\end{equation}
Additionally, the expression of $c_s^2$ \eqref{eq:cs2} also allows us to recast the perturbed energy density as follows
\begin{equation}
    \delta\rho=Y_0^{1-2\alpha}\Biggr[2\delta V+\frac{1}{c_s^2}(\delta X-\delta V)\Biggr].
\end{equation}
Lastly, taking into account the relation \eqref{eq:Hk} and combining equations \eqref{eq:deltaV} and \eqref{eq:rhophi0}, the above expression yields:
\begin{equation}
    \delta\rho=-\frac{3\Hb}{\phi_0'}\left(\rho_0+p_0\right)\delta\phi+\frac{H_K(Y_0)}{c_s^2}\left(\delta X-\frac{X_0'}{\phi_0'}\delta\phi\right).
\end{equation}
This last expression can be once more simplified by considering equations \eqref{eq:sum rhop} and \eqref{eq:deltaX} along with recasting $X_0'$ in terms of derivatives of $\phi_0$ via equation \eqref{eq:X0}. Following this procedure, we obtain the final expression for the energy density perturbation:
\begin{equation}
    \delta\rho = \left(\rho_0+p_0\right) \left\{ -3\Hb\frac{\delta\phi}{\phi_0'}+\frac{1}{c_s^2}\left[\left(\frac{\delta\phi}{\phi_0'}\right)'+\Hb\frac{\delta\phi}{\phi_0'}-\Phi\right]\right\}.
    \label{eq:deltarho fin}
\end{equation}

In the case of the component $\delta\tensor{T}{^0_i}$ \eqref{eq:deltaT0i}, the expression \eqref{eq:sum rhop} can be used, so that
\begin{equation}
    \delta\tensor{T}{^0_i} = \partial_i\left[\left(\rho_0+p_0\right)\frac{\delta\phi}{\phi_0'}\right],
    \label{eq:deltaT0i fin}
\end{equation}
where we have taken into account that $\rho_0$ and $p_0$, as background functions, lack of spatial derivatives.
Furthermore, we turn our attention to the Einstein equations \eqref{eq:G00}--\eqref{eq:Gij}. Since we have $\delta\tensor{T}{^i_j}\propto\tensor{\delta}{^i_j}$ \eqref{eq:deltaTij}, the $i\neq j$ equations easily imply
\begin{equation}
    \partial_i\partial_jD=0\quad\Rightarrow\quad D=0,
    \label{eq:Dnull}
\end{equation}
after removing the background functions. 

From now on, we choose to work in the longitudinal gauge, that is $E=B=0$. Thus, taking into account equation \eqref{eq:Dnull} in definition \eqref{eq:Ddef}, we get the following relation between the scalar potentials
\begin{equation}
 \Psi=\Phi.
\end{equation}
Thus, the Einstein equations \eqref{eq:G00} and \eqref{eq:G0i} now read:
\begin{align}
    4\pi Ga^2\delta\rho & = \nabla^2\Phi-3\Hb\left(\Phi'+\Hb\Phi\right),
    \label{eq:deltaG00}
    \\
    4\pi Ga^2\delta\tensor{T}{^0_i} & = \partial_i\left(\Phi'+\Hb\Phi\right).
    \label{eq:deltaG0i}
\end{align}
Combining equations \eqref{eq:deltarho fin} and \eqref{eq:deltaG00} together, along with the background equations \eqref{eq:FLRW conf} and \eqref{eq:FLRW and cons conf}, we obtain 
\begin{equation}
    \nabla^2\Phi = \frac{4\pi Ga^2\left(\rho_0+p_0\right)}{c_s^2\Hb}\left(\Hb\frac{\delta\phi}{\phi_0'}+\Phi\right)';
    \label{eq:nablaphi}
\end{equation}
whereas the combination of equations \eqref{eq:deltaT0i fin} and \eqref{eq:deltaG0i} yields
\begin{equation}
    \left(\frac{a^2\Phi}{\Hb}\right)' = \frac{4\pi Ga^4\left(\rho_0+p_0\right)}{\Hb^2}\left(\Hb\frac{\delta\phi}{\phi_0'}+\Phi\right),
    \label{eq:phiH}
\end{equation}
%\fr
Following \cite{Mukhanov:2005sc}, let us introduce the Mukhanov variables:
\begin{align}
    u & =  \frac{1}{4\pi G}\frac{\Phi}{\sqrt{\rho_0+p_0}},
    \label{eq:u Mukhanov}
    \\
    v & =  \sqrt{\frac{\rho_0+p_0}{2X_0c_s^2}}a\left(\delta\phi+\frac{\phi_0'}{\Hb}\Phi\right).
    \label{eq:v Mukhanov}
\end{align}
This choice transforms the previous equations \eqref{eq:nablaphi} and \eqref{eq:phiH} into
\begin{align}
    c_s\nabla^2u & =  z\left(\frac{v}{z}\right)';
    \label{eq:nablau Mukhanov}
    \\
    c_s v & =  \theta\left(\frac{u}{\theta}\right)';
    \label{eq:csv Mukhanov}
\end{align}
%º
where we have defined the following quantities
\begin{align}
    z & =   \frac{a^2\sqrt{\rho_0+p_0}}{c_s\Hb},
    \label{eq:z Mukhanov}
    \\
    \theta & =  \frac{1}{c_sz}.
    \label{eq:theta Mukhanov}
\end{align}
Substituting then expression \eqref{eq:csv Mukhanov} into equation \eqref{eq:nablau Mukhanov} and rearranging, we find the following closed PDE for $u$:
\begin{equation}
    u''-c_s^2\nabla^2 u-\frac{\theta''}{\theta}u=0.
    \label{eq:PDE u}
\end{equation}
Essentially, given the solutions for $u$ of this last PDE, we will be able to obtain the remaining quantities, namely
\begin{align}
    \Phi = & 4\pi G\sqrt{\rho_0+p_0}u,
    \\
    \delta\phi = & \frac{\phi_0'(a\Phi)'}{4\pi G a^3\left(\rho_0+p_0\right)};
\end{align}
where we have used  expressions \eqref{eq:u Mukhanov} and \eqref{eq:deltaG0i}, respectively. Note that, formally, the results are equivalent to those of the Diff theory, but we should recall that the symmetry breaking lies in the matter sector. Thus, the new physical information is contained in the energy density $\rho_0$, the pressure $p_0$ and the speed of sound $c_s$.
In next section we will address the quantization of these variables with the goal of finding the TDiff primordial power-spectrum.
\subsection{Quantization}\label{sec:QUANTIZATION}
In order to canonically quantize the introduced variables from the previous section, we should find the corresponding action for the cosmological perturbations. Instead of directly expanding the action for the gravitational and scalar fields to second order in perturbations, we can follow \cite{Mukhanov:2005sc} so that the required action can be deduced directly from the EoMs \eqref{eq:nablau Mukhanov} and \eqref{eq:csv Mukhanov} up to a time-independent operator $\hat{O}$. That is, we consider the following action 
\begin{equation}
    \begin{split}
         S^{(2)} & = \int d^4x \bigg[\left(\frac{v}{z}\right)'\hat O\left(\frac{u}{\theta}\right)\\
         & \,\,\,\,\,\,-\frac{c_s^2}{2}(\nabla^2u)\hat O(u)+\frac{c_s^2}{2} v\hat O(v)\bigg].
    \end{split}
    \label{eq:S2 pert v1}
\end{equation}
This expression can be further simplified, by using the EoM \eqref{eq:nablau Mukhanov}, as follows
\begin{equation}
    S^{(2)} = \int d^4x \frac{1}{2} \left[z^2\left(\frac{v}{z}\right)'\frac{\hat O}{\nabla^2}\left(\frac{v}{z}\right)'+c_s^2v\hat O(v)\right].
    \label{eq:S2 pert v2}
\end{equation}
The operator $\hat O$ can be determined if we compare the above expression with the action \eqref{eq:SphiDiff}, considering the de Sitter limit, i.e. $\frac{\phi_0'}{\Hb}\to0$, along with the massless limit, i.e., $V''(\phi_0)\to0$. The resulting action is
\begin{equation}
    S^{(2)} = \int d^4x a^2H_K(Y_0) \left[\left(\delta\phi'\right)^2-(\vec\nabla\delta\phi)^2\right].\label{eq:S2H}
\end{equation}
We can now apply these limits to the variables and see from \eqref{eq:v Mukhanov} that $v\to\sqrt{H_K}a\delta\phi$, as $c_s^2\to1$. Thus, actions \eqref{eq:S2 pert v2} and \eqref{eq:S2H} shall be equal if we take $\hat O=\nabla^2$. In that case, the action \eqref{eq:S2 pert v1} yields
\begin{equation}
    S^{(2)} = \int d^4x \frac{1}{2} \left(v'^2+c_s^2v\nabla^2v+\frac{z''}{z}v^2\right).
    \label{eq:S2 pert v3}
\end{equation}
Indeed, varying the action with respect to $v$, we readily obtain the sought-after EoM
\begin{equation}
    v''-c_s^2\nabla^2 v-\frac{z''}{z}v=0,
    \label{eq:PDE v}
\end{equation}
which is the so-called Mukhanov-Sasaki equation. It formally takes the same form as in the Diff case. In \ref{sec:ApA},  we include a discussion on the number of propagating degrees of freedom present in the theory and an alternative derivation of the quadratic action \eqref{eq:S2 pert v3}.

In order to obtain $c_s^2$ appearing in \eqref{eq:PDE v} in  terms of the slow-roll parameters, we can evaluate the expression \eqref{eq:eps slr cov} on the background, so that
\begin{equation}
    \frac{V_0}{X_0}\simeq\frac{3}{2\alpha\varepsilon}.
    \label{eq:V0X0}
\end{equation}
and substituting in \eqref{eq:cs2} we obtain\footnote{Notice that the speed of sound can also be obtained as $c_s^2=\delta p/\delta \rho$ evaluated in the rest gauge in which $\delta \phi=0$.}
\begin{equation}
    c_s^2\simeq1+\frac{2}{3}(1-2\alpha)\varepsilon.
    \label{eq:cs2 fin}
\end{equation}
Therefore, the speed of sound shall be nearly constant in the slow-roll regime and, as expected, we recover $c_s^2=1$ in the Diff limit.

In addition, we can explicitly recast the quantity $z$ in terms of the slow-roll parameters. To do so, equation \eqref{eq:sum rhop} can be substituted into equation \eqref{eq:z Mukhanov}, so that 
\begin{equation}
    z=\frac{a\phi_0'\sqrt{H_K(Y_0)}}{c_s\Hb}.
    \label{eq:z fin}
\end{equation}
We can compute the first derivative and find
\begin{equation}
    \frac{z'}{z}=\Hb+\frac{\phi_0''}{\phi_0'}+\frac{\left[{(Y_0^{1-2\alpha})^\frac{1}{2}}\right]'}{(Y_0^{1-2\alpha})^\frac{1}{2}}-\frac{\Hb'}{\Hb},
    \label{eq:z'}
\end{equation}
where we have used equation \eqref{eq:Hk plaw}. If we now recall the expression \eqref{eq:dep Y-a}, rightly expressed in conformal time, we are able to recast the above equation as follows
\begin{equation}
    \frac{z'}{z}\simeq\left[1+2\alpha\varepsilon-\delta\right]\Hb,
    \label{eq:z' fin}
\end{equation}
where we bear in mind expression \eqref{eq:eps conformal time} for $\varepsilon$ and, in addition, that the definition of $\delta$ in \eqref{eq:Diff eta} yields in conformal time \eqref{eq:conformal time}
\begin{equation}
    \delta=1-\frac{\phi_0''}{\phi_0'\Hb}.
    \label{eq:delta conformal time}
\end{equation}

Making use of the fact that $\varepsilon'\sim\mathcal{O}(\varepsilon^2)$ and $\delta'\sim\mathcal{O}(\varepsilon^2)$ \cite{Liddle:1994dx}, we can use \eqref{eq:z' fin} to find
\begin{equation}
\begin{split}
     \frac{z''}{z} & \simeq\left[2+(6\alpha-1)\varepsilon-3\delta\right]\Hb^2\simeq\\
     & \simeq\frac{1}{\tau^2}\left[2+3(2\alpha+1)\varepsilon-3\delta\right],
\end{split}
\label{eq:z'' fin}
\end{equation}
where in the last step we have used equation \eqref{eq:Hetaeps}.

In light of these results, the Mukhanov-Sasaki equation \eqref{eq:PDE v} in Fourier space reads
\begin{equation}
        v_k''-\left[k^2c_s^2-\frac{1}{\tau^2}\left(\nu^2-\frac{1}{4}\right)\right]v_k=0,
    \label{eq:PDE v Bessel}
\end{equation}
where we have defined
\begin{equation}
    \nu\simeq\frac{3}{2}+(2\alpha+1)\varepsilon-\delta.
    \label{eq:nu}
\end{equation}
This equation turns into a canonical Bessel equation \cite{Baumann:2022mni}, whose general solution is just
\begin{equation}
    \begin{split}
        v_k(\tau) & = C_1(k)(-\tau)^{\frac{1}{2}}H_\nu^{(1)}(-c_sk\tau) \\
        &\,\,\,\,\,\,+C_2(k)(-\tau)^{\frac{1}{2}}H_\nu^{(2)}(-c_sk\tau).
        \label{eq:vk sol}
    \end{split}
\end{equation}
Here, $H_\nu^{(1,2)}$ are the Hankel functions of the first and second kind,  and $C_{1,2}(k)$ are two integration constants. Furthermore, the solution for positive frequency modes in the sub-Hubble regime $|c_sk\tau|\gg1$ can be only achieved by the asymptotic behavior of $H_\nu^{(1)}$. Thus, a suitable choice of the constants yields, for  the Bunch-Davies vacuum, the following form for the solutions
\begin{equation}
    v_k(\tau)=\frac{1}{2}\sqrt{\frac{\pi}{c_sk\bar V}}e^{i\left(\nu+\frac{1}{2}\right)\frac{\pi}{2}}\sqrt{-c_sk\tau}H_\nu^{(1)}(-c_sk\tau).
    \label{eq:vk subH}
\end{equation}
with $\bar V$ the finite spatial volume. 
In the super-Hubble regime $|c_sk\tau|\ll1$, the solution reads
%º
\begin{equation}
    v_k(\tau)=\frac{2^{\nu-\frac{3}{2}}}{{\sqrt{2c_sk\bar V}}}\frac{\Gamma(\nu)}{\Gamma\left(\frac{3}{2}\right)}\left(-c_sk\tau\right)^{\frac{1}{2}-\nu}.
    \label{eq:vk superH}
\end{equation}

The primordial power-spectrum for the curvature perturbation $\zeta$ can be now obtained. For this purpose, we can combine the variables \eqref{eq:v Mukhanov} and \eqref{eq:z Mukhanov}, so that
\begin{equation}
    \frac{v}{z}=\Hb\frac{\delta\phi}{\phi_0'}+\Phi=\zeta.
    \label{eq:curv pert}
\end{equation}
Using this last relation, we can write
\begin{equation}
    P_\zeta(\tau,k)=\frac{k^3\bar V}{2\pi^2}|\zeta_k|^2=\frac{k^3\bar V}{2\pi^2z^2}|v_k|^2.
\end{equation}
In order to simplify the above expression, we can take into account equations \eqref{eq:eps conformal time}, \eqref{eq:FLRW and cons conf} and \eqref{eq:sum rhop}, so that
\begin{equation}
    \varepsilon=\frac{\Hb^2-\Hb'}{\Hb^2}=\frac{4\pi G}{\Hb^2}\phi_0'^2H_K(Y_0),
    \label{eq:eps conf}
\end{equation}
which can be substituted back into equation \eqref{eq:z Mukhanov} and find
\begin{equation}
    \frac{1}{z^2}=\frac{4\pi Gc_s^2}{a^2\varepsilon}.
    \label{eq:z2 inv}
\end{equation}
Finally, to leading order in the slow-roll approximation we get
\begin{equation}
     P_\zeta(\tau,k)\simeq\frac{4\pi G}{\varepsilon}\left(\frac{H}{2\pi}\right)^2\left(\frac{k}{aH}\right)^{3-2\nu},
     \label{eq:Pdseta}
\end{equation}
where we have used that $c_s^2\simeq1$ from \eqref{eq:cs2 fin}. Although \eqref{eq:Pdseta} is the same formal expression as the one that can be obtained in the Diff case, the background evolution (and, therefore, $H$) is now affected by the TDiff function. More importantly, it has to be stressed that the parameter $\alpha$ appears now in the definition of $\nu$ \eqref{eq:nu}. Equivalently, we can work with the spectral index using equation \eqref{eq:nu}
\begin{equation}
    n_S=4-2\nu\simeq1-2(2\alpha+1)\varepsilon+2\delta
    \label{eq:ns}
\end{equation}
which depends on the TDiff parameter $\alpha$, and the scalar amplitude at a given pivot-scale $k_\ast$ 
\begin{equation}
    A_S=\frac{4\pi G}{\varepsilon}\left.\left(\frac{H}{2\pi}\right)^2\right\rvert_{k_\ast=aH}.
    \label{eq:As}
\end{equation}
Thus, the primordial power-spectrum just becomes
\begin{equation}
     P_\zeta(\tau,k)=A_S\left(\frac{k}{k_\ast}\right)^{n_S-1}.
\end{equation}
Furthermore, the combination of the Friedmann equation \eqref{eq:FLRW} and the energy density \eqref{eq:rho TDiff plaw} in the slow-roll approximation allows us to write
\begin{equation}
    H^2\simeq\frac{8\pi G}{3}Y^{1-2\alpha}2\alpha V(\phi),
    \label{eq:H2 simp}
\end{equation}
so that, taking into account equation \eqref{eq:prim par slr}, the scalar amplitude can be rewritten as
\begin{equation}
    A_S=(2\alpha)^3Y^{2(1-2\alpha)}\frac{128\pi}{3}\left.\frac{G^3V^3}{V'^2}\right\rvert_{k_\ast=aH},
    \label{eq:As simp}
\end{equation}
 Once again, note that we successfully recover the Diff expression for the scalar amplitude if we make $\alpha=1/2$.

For later convenience, we can recast the TDiff function $Y$ in terms of the potential $V(\phi)$ and the number of $e$-folds. Let us define the integral that appears in the expression fo the number of $e$-folds during slow-roll, equation \eqref{eq:TDiff efolds}, as
\begin{equation}
    I_\alpha(\phi_\ast)=\int^{\phi_f}_{\phi_\ast}d\phi \frac{V^{\frac{1}{2\alpha}}}{V'}.
    \label{eq:Ia}
\end{equation}
Additionally, we are able to relate the value of the field at the end of inflation $Y_f$ with that one evaluated at the time when the pivot-scale left the horizon $Y_\ast$ via the expression \eqref{eq: dep YV chain}. In doing so, we can substitute the result back into \eqref{eq:TDiff efolds} and find
\begin{equation}
    Y_\ast^{1-2\alpha}=\left.-\frac{N}{16\pi G \alpha I_\alpha} V^{\frac{1}{2\alpha}-1}\right\rvert_{k_\ast=aH}.
    \label{eq:YVN}
\end{equation}
Note that this quantity is always positive for those models with $\phi_\ast>\phi_f$, i. e., the field is rolling down from right to left. Therefore, the scalar amplitude \eqref{eq:As simp} takes the final form of
\begin{equation}
    A_S=\left.\frac{4\alpha G}{3\pi}\left(\frac{N}{I_\alpha}\right)^2\frac{V^{\frac{1}{\alpha}+1}}{V'^2}\right\rvert_{k_\ast=aH},
    \label{eq:As N}
\end{equation}
which essentially only depends on the considered number of $e$-folds and the potential form. We are now ready to compare the predictions of the TDiff models with observational data.

\section{PHENOMENOLOGY OF TDIFF SLOW-ROLL}\label{sec:phenomenology}
In this section, we will focus on power-law potentials and write the expressions for the relevant observables. Then, we will compare the predictions of TDiff inflation  with the available observational data.
\subsection{Power-law potentials}\label{sec:power-law}
Let us now consider throughout this section the potential function of a general power-law of the form
\begin{equation}
    V(\phi)=\lambda\phi^p,
    \label{eq:V plaw}
\end{equation}
with $p>0$. It should be noted that models with inverse power laws require an extra mechanism to end inflation \cite{Martin:2013tda}. The slow-roll parameters in the slow-roll approximation, given by equations \eqref{eq:prim par slr} and \eqref{eq:sec par slr}, can then be written as 
\begin{align}
\varepsilon & \simeq \frac{p^2}{64\pi G\alpha^2}\frac{Y^{2\alpha-1}}{\phi^2}, \label{eq:eps plaw}
        \\
        \eta & \simeq\frac{p(p-1)}{16\pi G\alpha}\frac{Y^{2\alpha-1}}{\phi^2}.\label{eq:eta plaw}
\end{align}
Thus, we can write
\begin{equation}
    \varepsilon\simeq\frac{\alpha_p}{\alpha}\eta,
    \label{eq:epseta plaw}
\end{equation}
where we have defined
\begin{equation}
    \alpha_p=\frac{p}{4(p-1)}.
    \label{eq:alpha pivot}
\end{equation}
In general, for those power values that satisfy $p>1$, we would have a positive $\alpha_p$, otherwise, for $p\in(0,1)$ we would have a negative value. Note also that $\eta<0$ for $p\in(0,1)$. From \eqref{eq:epseta plaw}, we see that the TDiff parameter $\alpha_p$ then serves as a pivot for the dominance of $\varepsilon$, i.e. $\varepsilon>|\eta|$ for $\alpha<|\alpha_p|$ or, on the contrary, $\varepsilon < |\eta|$ for $\alpha>|\alpha_p|$. We remark that we are excluding the case $p=1$, as $\eta\propto V''$ vanishes and $\varepsilon$ dominates accordingly. Naturally, for $\alpha=\alpha_p$ both slow-roll parameters agree. 
\begin{figure}[t]
    \centering
    \includegraphics[width=\linewidth]{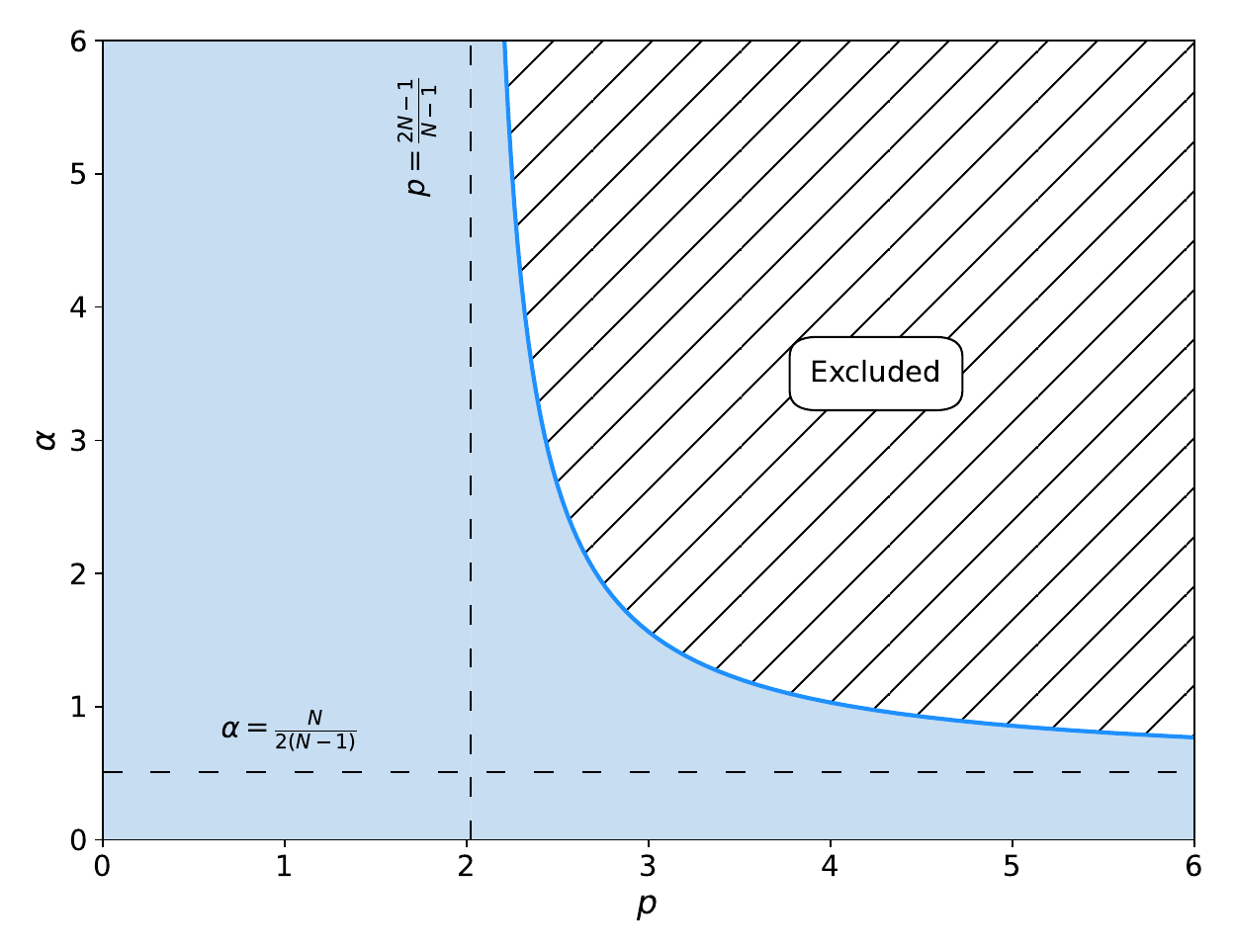}
    \caption{Excluded and allowed values of $\alpha$ as a function of the power $p$ for a given number of $e$-folds $N$. The striped area represents the excluded $\alpha$ values, which corresponds to $N>N_{\rm max}$. The colored region represents the allowed $\alpha$ values. The blue colored line is the border of the region given by \eqref{eq:condition alpha}, presenting a vertical asymptote at $p=(2N-1)/(N-1)$ and a horizontal asymptote at $\alpha=N/[2(N-1)]$.}
    \label{fig:cond alp}
\end{figure}
Applying now the condition for the end of slow-roll, $\text{max}\{\varepsilon(\phi_f),|\eta(\phi_f)|\}=1$ in equations \eqref{eq:eps plaw} and \eqref{eq:eta plaw}, we find the value of the field
\begin{subequations}\label{eq:phif plaw}
\begin{empheq}[left={\phi_f^2 = \empheqlbrace}]{align}
    \frac{p^2}{64\pi G\alpha^2}Y_f^{2\alpha-1} &,\quad  \alpha<|\alpha_p| \label{eq:phif eta<eps}\\
    \frac{p|p-1|}{16\pi G\alpha}Y_f^{2\alpha-1} &,\quad \alpha>|\alpha_p| \label{eq:phif eta>eps}
\end{empheq}
\end{subequations}
In order to express these quantities in terms of the number of $e$-folds, we can perform the integral \eqref{eq:Ia} for the power-law potential \eqref{eq:V plaw}; this yields
\begin{equation}
I_\alpha(\phi_*)=\frac{\lambda^{\frac{1}{2\alpha}-1}}{p\,\sigma_p}\left(\phi_f^{\sigma_p}-\phi_*^{\sigma_p}\right),
\label{eq:Iplaw}
\end{equation}
where we have defined the function
\begin{equation}
    \sigma_p= p\left(\frac{1}{2\alpha}-1\right)+2.
    \label{eq:sigma p}
\end{equation}
By substituting $I_\alpha(\phi_*)$ and \eqref{eq:V plaw} into equation \eqref{eq:TDiff efolds}, one obtains
\begin{equation}
    N\simeq \frac{16\pi G\alpha}{p\,\sigma_p}Y_f^{1-2\alpha}\phi_f^2\left[\left(\frac{\phi_*}{\phi_f}\right)^{\sigma_p}-1\right].\label{eq:Npivot}
\end{equation}
Essentially, the above expression allows us to find the value of the field at the pivot-scale
\begin{equation}
    \phi_\ast=\phi_f\left(1+\frac{p\sigma_pNY_f^{2\alpha-1}}{16\pi G\alpha\phi_f^2}\right)^\frac{1}{\sigma_p}, 
    \label{eq:phi*}
\end{equation}
Note that for $\sigma_p>0$ we can always obtain $\phi_*>\phi_f$ for any value of $N$. However, for $\sigma_p<0$ this may not be the case and, therefore,  some restrictions are imposed on the allowed $\alpha$ values.
Indeed, in such case we may rewrite equation \eqref{eq:Npivot} as
\begin{equation}
    N\simeq N_{\rm max}\left[1-\left(\frac{\phi_f}{\phi_*}\right)^{|\sigma_p|}\right],\quad {\rm for}\,\,\sigma_p<0,\label{eq:Npivot0}
\end{equation}
with the definition of 
\begin{equation}\label{Nmax}
    N_{\rm max}= \frac{16\pi G\alpha}{p\,|\sigma_p|}Y_f^{1-2\alpha}\phi_f^2.
\end{equation}
Thus, in the limit $\phi_*\rightarrow\infty$,  we get $N\rightarrow N_{\rm max}$, so that $N_{\rm max}$ is indeed the maximum number of e-folds produced by the model with a particular $(p, \alpha)$. If we now substitute the field $\phi_f$ \eqref{eq:phif eta<eps} or \eqref{eq:phif eta>eps} into this last equation \eqref{Nmax}, we obtain
\begin{subequations}\label{eq:Nmax}
\begin{empheq}[left={ N_{\rm max} = \empheqlbrace}]{align}
   \frac{p}{4|\sigma_p|\alpha} &,\quad  \alpha<|\alpha_p| \label{eq:phiast eta<eps2}\\
   \frac{|p-1|}{|\sigma_p|} &,\quad \alpha>|\alpha_p|. \label{eq:phiast eta>eps2}
\end{empheq}
\end{subequations}
The physical solutions must satisfy the condition for the number of $e$-folds $N< N_{\rm max}$. Taking into account the form of $\sigma_p$ \eqref{eq:sigma p}, the following inequalities are obtained
\begin{subequations}\label{eq:cond phiast}
\begin{empheq}[left={\empheqlbrace}]{align}
    \frac{p}{4}\left(2+\frac{1}{N}\right)>(p-2)\alpha &,\quad  \alpha<|\alpha_p| \label{eq:cond phiast eta<eps}\\
   \frac{pN}{2}>-\left[|p-1|+(2-p)N\right]\alpha &,\quad \alpha>|\alpha_p| \label{eq:cond phiast eta>eps}
\end{empheq}
\end{subequations}
It can be seen that these conditions impose restrictions on $\alpha$ only for exponents values $p>\frac{2N-1}{N-1}$ provided $\sigma_p<0$. These restrictions can be written as
\begin{equation}
  \alpha< \frac{pN}{2\left[(p-2)N-|p-1|\right]}.
  \label{eq:condition alpha}
\end{equation}
In figure \ref{fig:cond alp} we have represented the above inequality. The colored  region shows the physically acceptable $\alpha$ values that satisfy $N<N_{\rm max}$, whereas the striped area represents the excluded values. As we can see, the limit $p\to\infty$ provides the lower bound $\alpha=N/[2(N-1)]$ and the limit $p\to(2N-1)/(N-1)$ yields $\alpha\to\infty$. Note that the condition $\sigma_p<0$ does not impose additional restrictions in this plot.

With all this information, the spectral index can be computed. Firstly, we can apply the definition \eqref{eq:ns}, evaluated at the pivot-scale $k_\ast$, and we should recall the relation \eqref{eq:epseta plaw} between the slow-roll parameters, so that
\begin{equation}
    n_S=1-4\left[1+\left(\frac{2}{p}-1\right)\alpha\right]\varepsilon_\ast.
    \label{eq:ns plaw}
\end{equation}
Secondly, the value of $\varepsilon_*$ is obtained via the substitution of equations \eqref{eq:YVN} and \eqref{eq:Iplaw} into equation \eqref{eq:eps plaw},
\begin{equation}
    \varepsilon_\ast=\frac{p}{4\alpha\sigma_pN}\left[1-\left(\frac{\phi_f}{\phi_\ast}\right)^{\sigma_p}\right].
    \label{eq:epsast}
\end{equation}
The remaining quotient can be computed by using \eqref{eq:phi*} combined with \eqref{eq:phif eta<eps} or \eqref{eq:phif eta>eps}, so that
\begin{subequations}\label{eq:phiast phif}
\begin{empheq}[left={\left(\frac{\phi_\ast}{\phi_f}\right)^{\sigma_p} = \empheqlbrace}]{align}
   1+\frac{4\alpha}{p}\sigma_pN &,\quad  \alpha<|\alpha_p| \label{eq:phiast eta<eps}\\
   1+\frac{\sigma_p}{|p-1|}N &,\quad \alpha>|\alpha_p| \label{eq:phiast eta>eps}
\end{empheq}
\end{subequations}
Combination of equation \eqref{eq:epsast} with equations \eqref{eq:phiast eta<eps} and \eqref{eq:phiast eta>eps} yields
\begin{subequations}\label{eq:epsast fin}
\begin{empheq}[left={\varepsilon_\ast= \empheqlbrace}]{align}
    \frac{p}{p+4\alpha\sigma_pN} &,\quad  \alpha<|\alpha_p| \label{eq:epsast eta<eps}\\
   \frac{p}{4\alpha(|p-1|+\sigma_pN)} &,\quad \alpha>|\alpha_p| \label{eq:epsast eta>eps}
\end{empheq}
\end{subequations}
Here, we should take into account the already discussed condition on $\alpha$ \eqref{eq:condition alpha}, otherwise the expressions may not be physically acceptable.

Lastly, we can substitute \eqref{eq:epsast eta<eps} or \eqref{eq:epsast eta>eps} into \eqref{eq:ns plaw} and write the final expression for the spectral index
\begin{subequations}\label{eq:ns fin}
\begin{empheq}[left={n_S-1= \empheqlbrace}]{align}
   -4p\frac{1+\left(\frac{2}{p}-1\right)\alpha}{p+4\alpha\sigma_pN} &,\quad  \alpha<|\alpha_p| \label{eq:ns fin eta<eps}\\
   -\frac{p}{\alpha}\frac{1+\left(\frac{2}{p}-1\right)\alpha}{|p-1|+\sigma_pN} &,\quad \alpha>|\alpha_p| \label{eq:ns fin eta>eps}
\end{empheq}
\end{subequations}

On the other hand, the tensor-to-scalar ratio can be likewise obtained, however, we shall remark a few nuances for this goal before. As we saw in section \ref{sec:TDIFF FRAMEWORK}, the total action is $S=S_{EH}+S^{\rm cov}_\phi$, which means that the gravitational sector does not feature directly the symmetry breaking down to TDiff. Thus, the Diff results for the tensor perturbations \cite{Mukhanov:2005sc,Baumann:2022mni} should still hold true. However, there is a stronger reason that supports this fact. Recalling the equation \eqref{eq:deltaTij} of the perturbed EMT, we point out the dependence $\delta\tensor{T}{^i_j} \propto\tensor{\delta}{^i_j}$, that is, the anisotropic stress certainly vanishes, since no residual contribution from the covariantizing field
$A_\mu$ enters the transverse-traceless sector at linear order. As a result, the transverse-traceless sector should then decouple from the matter perturbations exactly as in the standard Diff case.  
Consequently, we are able to write the power-spectrum as in the Diff theory
\begin{equation}
     P_T(\eta,k)=A_T\left(\frac{k}{k_\ast}\right)^{n_T},
\end{equation}
where now the amplitude is
\begin{equation}
     A_T=64\pi G\left.\left(\frac{H}{2\pi}\right)^2\right\rvert_{k_\ast=aH},
     \label{eq:AT}
\end{equation}
and the tensorial index reads
\begin{equation}
     n_T\simeq-2\varepsilon.
     \label{eq:nT}
\end{equation}
The tensor-to-scalar ratio is defined as the quotient between the scalar and tensorial amplitude at the same pivot-scale. Combining expressions \eqref{eq:As} and \eqref{eq:AT}, we can write:
\begin{subequations}\label{eq:r}
\begin{empheq}[left={r=\frac{A_T}{A_S}=16\varepsilon_\ast= \empheqlbrace}]{align}
    \frac{16p}{p+4\alpha\sigma_pN} &,\,\,  \alpha<|\alpha_p| \label{eq:r eta<eps}\\
    \frac{4p}{\alpha(|p-1|+\sigma_pN)} &,\,\, \alpha>|\alpha_p| \label{eq:r eta>eps}
\end{empheq}
\end{subequations}
where we have used \eqref{eq:epsast eta<eps} or \eqref{eq:epsast eta>eps}. Note that we still have the following consistency relation $r=-8\,n_T$. 

For later comparison with the experimental data, the curve $r=r(n_S)$ can be easily obtained with the direct combination of the ratio \eqref{eq:r eta<eps} or \eqref{eq:r eta>eps} with the corresponding spectral index \eqref{eq:ns fin eta<eps} or \eqref{eq:ns fin eta>eps}, respectively, so that
\begin{equation}
    r=\frac{4(1-n_S)}{1+\left(\frac{2}{p}-1\right)\alpha}.
    \label{eq:rns palpha}
\end{equation}
Notice that this expression holds for both cases,  $\alpha<|\alpha_p|$ and $\alpha>|\alpha_p|$. We can also see that the dependencies on $p$ and $\alpha$ are degenerate in \eqref{eq:rns palpha}.

Finally, we can study the following limiting cases.  On the one hand, we have the limit $\alpha\to0$, where $|\alpha_p|>0$ \eqref{eq:alpha pivot} (as we are excluding $p=1$), so that we can apply equation \eqref{eq:ns fin eta<eps} and find
\begin{equation}
    \underset{\alpha\to0}{\rm lim}\,n_S-1 = -\frac{4}{1+2N}
    \label{eq:ns lim0}
\end{equation}
and from \eqref{eq:r eta<eps}
\begin{equation}
    \underset{\alpha\to0}{\rm lim}\,r = \frac{16}{1+2N}.
    \label{eq:r lim0}
\end{equation}
On the other hand, the limit $\alpha\rightarrow\infty$ only exists for $\alpha>|\alpha_p|$. In such case, we have to take into account the condition \eqref{eq:condition alpha}. For power values $p<(2N-1)/(N-1)$, the limit is physically valid, so the spectral-index reads
\begin{equation}
    \underset{\alpha\to\infty}{\rm lim}\,n_S-1 = -\frac{2-p}{|p-1|+(2-p)N}
    \label{eq:ns lim inf}
\end{equation}
and one gets for the tensor to scalar ratio 
\begin{equation}
    \underset{\alpha\to\infty}{\text{lim}}\, r =0,
    \label{eq:r lim inf}
\end{equation}
Once we have derived the spectral index and the tensor to scalar ratio, we are finally able to compare the TDiff results to the available experimental constraints in the $n_S$--$r$ plane. The next subsection will be devoted to this analysis.
\begin{figure}[t]
    \centering
    \includegraphics[width=\linewidth]{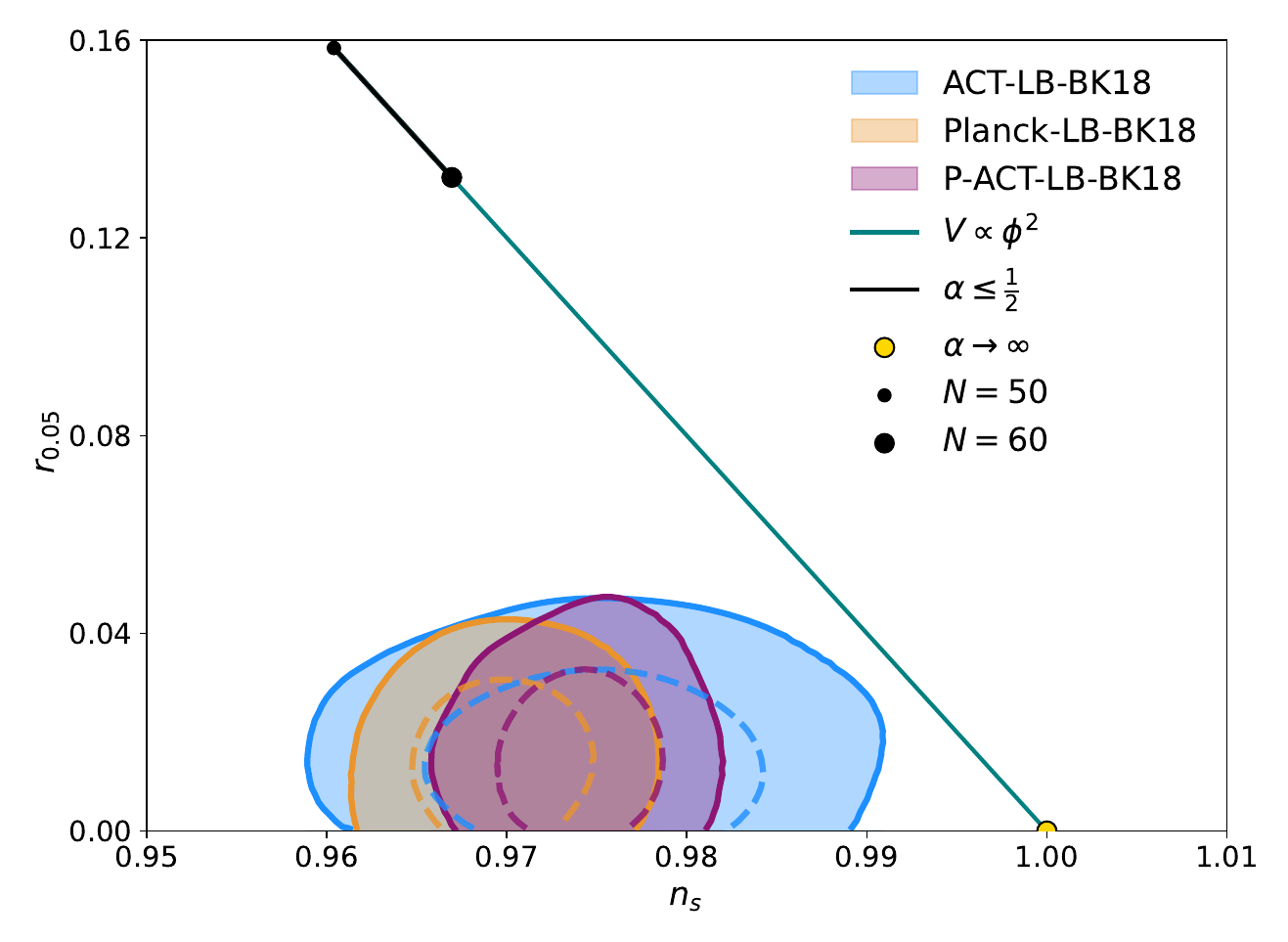}
    \caption{Constraints on the scalar and tensor primordial spectra at $k_\ast=0.05 \,\,\text{Mpc}^{-1}$ for 95\% CL region (solid line) and 68\% CL (dashed line), represented in the $n_S$--$r$ plane. The panel combines datasets from ACT (blue), re-run data of \textit{Planck 2018} with Sroll2 dataset (orange) and P-ACT (purple). All datasets make use of DESI Data Release 1 (DR1) and the contours for DESI DR2 have not been included, as the changes for P-ACT are hardly noticeable. In all cases, the dataset includes measurements of CMB lensing, BAO (LB) and CMB B-modes of polarization (BK18). The quadratic potential has been represented along variations of the TDiff parameter $\alpha$ for the number of $e$-folds of inflation $N\in[50,60]$. The Diff case is represented in black and overlaps with values of $\alpha\leq1/2$. The teal colored line represents values of $\alpha>\frac{1}{2}$.}
    \label{fig:ACT data V2}
\end{figure}
\begin{figure*}[t]
    \centering
    \includegraphics[width=\linewidth]{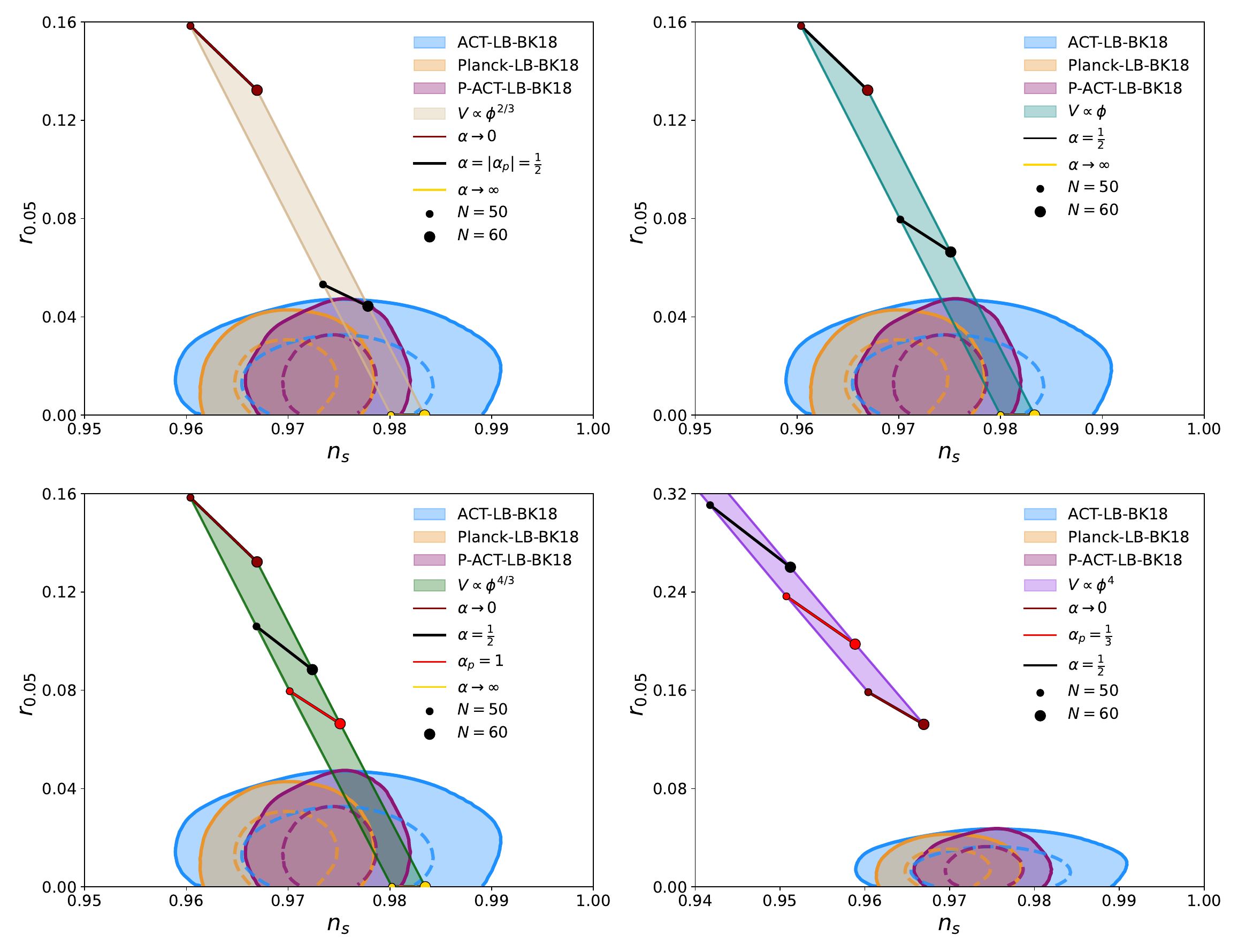}
    \caption{Constraints on the scalar and tensor primordial spectra at $k_\ast=0.05 \,\,\text{Mpc}^{-1}$ for 95\% CL region (solid line) and 68\% CL (dashed line), represented in the $n_S$--$r$ plane. The various colored bands show examples of power-law potentials along variations of the TDiff parameter $\alpha$ for the number of $e$-folds of inflation $N\in[50,60]$. The black colored lines correspond in every panel to the Diff case. The curves with $\alpha_p$ correspond to the transition value which indicates the change of the dominance for the slow-roll parameters. The $V\propto\phi$ plot lacks this line, as $\varepsilon$ always dominates.}
    \label{fig:ACT data Vs}
\end{figure*}
\subsection{Results comparison}\label{sec:Results comparison}
We are interested in comparing the predictions of the TDiff models  with data from  \textit{Planck Collaboration} \cite{Planck:2018jri} and \textit{Atacama Cosmology Telescope} (ACT) \cite{ACT:2025tim} observations. In particular, we will confront the previous results for $(n_S,r)$ with the confidence regions in reference \cite{ACT:2025tim} where  the tensor-to-scalar ratio is measured at the pivot-scale $k_\ast=0.05 \,\,\text{Mpc}^{-1}$, hereinafter referred as $r_{0.05}$. The results are shown in figures \ref{fig:ACT data V2} and \ref{fig:ACT data Vs} for different power-law potentials.

Firstly, the quadratic potential case $\phi^2$ has been represented in figure \ref{fig:ACT data V2} for  $\alpha>0$. In the present case, the substitution of $p=2$ into equation \eqref{eq:rns palpha} yields the curve $r=4(1-n_S)$, which is the same obtained in the Diff case. Nevertheless, we now have for the spectral index the following expressions
\begin{subequations}\label{eq:ns phi2}
\begin{empheq}[left={n_S-1= \empheqlbrace}]{align}
   -\frac{2}{N+\frac{1}{2}} &,\quad  \alpha<1/2 \label{eq:ns phi2 eta<eps}\\
   -\frac{2}{N+\alpha} &,\quad \alpha>1/2  \label{eq:ns phi2 eta>eps}
\end{empheq}
\end{subequations}
and, for the tensor to scalar ratio,
\begin{subequations}\label{eq:r phi2}
\begin{empheq}[left={r= \empheqlbrace}]{align}
    \frac{8}{N+\frac{1}{2}} &,\quad  \alpha<1/2 \label{eq:r phi2 eta<eps}\\
    \frac{8}{N+\alpha} &,\quad\alpha>1/2 \label{eq:r phi2 eta>eps}
\end{empheq}
\end{subequations}
where now $\alpha_p=1/2$ \eqref{eq:alpha pivot}. As we can see, for $\alpha<1/2$ we obtain the same expressions as in the Diff case, whereas for $\alpha>1/2$ the TDiff couplings could suppress the values of $n_S-1$ and $r$. Note that the number of $e$-folds is degenerate in this last case with $\alpha$.
Despite this new TDiff phenomenology and even though the ACT dataset has found constraints slightly closer to $n_S=1$, the quadratic potential still remains disfavored by the experimental data, as shown in figure \ref{fig:ACT data V2}. Although for larger values of $\alpha$ this potential can lead to smaller values of $r_{0.05}$ than the general relativistic model with similar number of $e$-folds, it predicts a value of $n_s$ which is still too large.

We can now turn our attention to figure \ref{fig:ACT data Vs}. Each panel shows a different power-law potential for different values of $\alpha$, including the limits \eqref{eq:ns lim0}--\eqref{eq:r lim inf} and the transition line corresponding to $\alpha_p$ \eqref{eq:alpha pivot}. In comparison to the quadratic case, which was degenerate in a single line, the TDiff regions are now wider.
In fact, ACT data disfavors power-laws with values $p>1$ in the Diff case; however the TDiff region is compatible with the $1\sigma$ regions for potential exponents $p<2$. 

For the sake of completeness, we have also analyzed the rest of possible cases with $p>(2N-1)/(N-1)$, taking into account the condition \eqref{eq:condition alpha}. As described before, the behavior is rather different, since $r$ has now a lower bound. The tensor to scalar ratio can grow until the limit \eqref{eq:condition alpha}, from which the results are not physically valid ($N>N_{\rm max}$). We have represented as an example the $\phi^4$ potential in the bottom right panel of figure \ref{fig:ACT data Vs}. As it happens in the Diff case, this potential is certainly disfavored, even though for $\alpha<1/2$ the tensor to scalar ratio is reduced and the spectral index increases.

\begin{figure}[t]
    \centering
    \includegraphics[width=\linewidth]{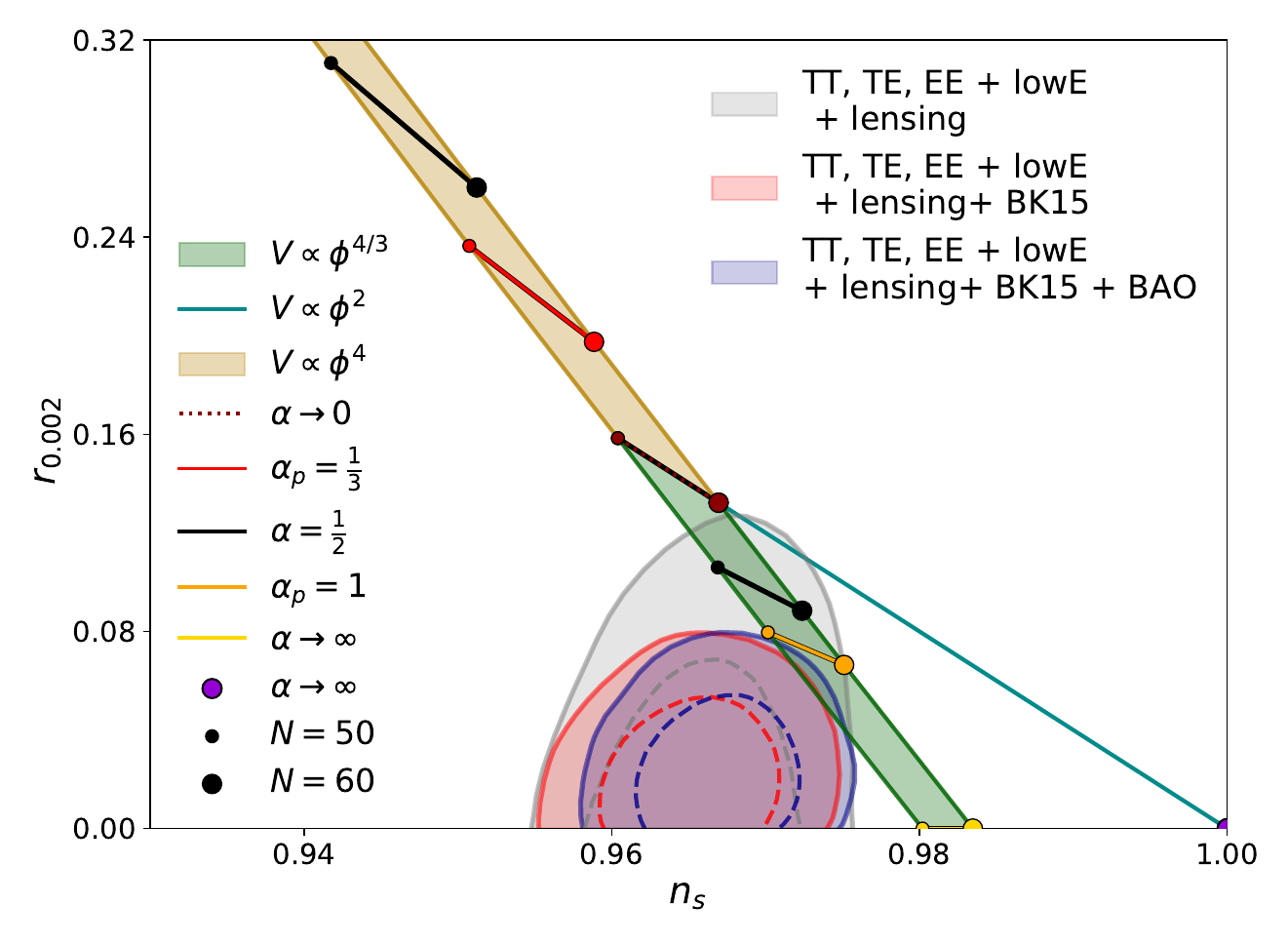}
    \caption{Constraints on the scalar and tensor primordial spectra at $k_\ast=0.002 \,\,\text{Mpc}^{-1}$ for 95\% CL region (solid line) and 68\% CL (dashed line), represented in the $n_S$--$r$ plane. The panels combine datasets from \textit{Planck 2018}: temperature--temperature power spectra (TT), temperature-E mode polarization power spectra (TE) and polarization-polarization power spectra (EE); low multipole E-mode polarization (lowE), CMB lensing (lensing), B-mode polarization (BK15), from BICEP-Keck 2015; and BAO. The color bands represent power-law potentials along variations of the TDiff parameter $\alpha$ for the number of $e$-folds of inflation $N\in[50,60]$. The Diff case is represented in black. The quadratic potential behavior for $\alpha\leq1/2$ overlaps with the region $\alpha\to0$ (brown colored dotted line) of the rest of represented potentials; for $\alpha>1/2$ it decreases following the teal colored line.}
    \label{fig:Planck data V2}
\end{figure}

Before continuing, there has been some noteworthy discussion about the ACT experimental data. Specifically, the ACT dataset is combined with information extracted from CMB lensing along with baryon acoustic oscillations (BAO, DESI Year-1), which is denoted as a whole by LB, and CMB B-modes (BK18). Nevertheless, the recent second data release of DESI (Year-3) \cite{DESI:2025zgx} has pointed out the rising tension between BAO and CMB data within $\Lambda$CDM. This could stand as a problem for the ACT confidence regions, as the dataset combines CMB data with DESI observations despite the increasing tension. In light of this, it is interesting to confront the TDiff model with CMB data alone. With that purpose, we have selected the \textit{Planck 2018} dataset \cite{Planck:2018jri,Demirel2025PlanckConstraints}. The analysis of the constraints in the $n_S$--$r$ parameter space for three potentials of interest can be seen in figure \ref{fig:Planck data V2}. Note that the pivot-scale is now $k_\ast=0.002\,\,\text{Mpc}^{-1}$. The corresponding constraints on the spectral index $n_S$ appear to be slightly lower, compared to ACT, whereas they are significantly greater for the tensor to scalar ratio $r$. As we can see, the TDiff models do not cross the $1\sigma$ region, unlike in figure \ref{fig:ACT data Vs}. However, we still find that potentials with $p<2$ are favored with respect to  potentials with larger exponents. In addition, the quadratic potential predictions are slightly improved for certain values of $\alpha>1/2$. We stress that the region $\alpha\to0$, corresponding to equations \eqref{eq:ns lim0} and \eqref{eq:r lim0} for all $p$, overlaps with the quadratic potential line for $\alpha\leq1/2$. As a brief reminder, these values are located on the line \eqref{eq:rns palpha}. 
\section{DYNAMICAL SYSTEM ANALYSIS}\label{sec:TDIFF QUADRATIC POTENTIAL}
In the last sections we have derived the TDiff primordial power-spectrum and also studied the compatibility of the datasets with the prediction of the TDiff models. So far we have concentrated ourselves in the slow-roll approximation, being the post-inflationary dynamics completely unexplored. The typical Diff models often feature oscillations of the inflaton around the minimum of the potential after the slow-roll regime which are related to the post-inflationary reheating phase. However, as we have seen, the TDiff theories showcase non-trivial behaviors, especially, the constraint \eqref{eq:Tcov const} that fixes the new physical degree of freedom $Y$. Therefore, we cannot expect a standard evolution for the inflaton after the end of slow-roll. 
In particular, the constraint plays a major role in the analysis because it drastically modifies the dynamics. The substitution of equation \eqref{eq:Hk plaw} into \eqref{eq:Tcov const} yields the following form:
\begin{equation}
(1-2\alpha)Y^{-2\alpha} \left[\frac{1}{2}\dot\phi^2-V(\phi)\right]=-\frac{c_\rho}{2}\label{eq:TDiff const ps}
\end{equation}
where $c_\rho=\text{const.}$ Note here that the quantity within brackets on the left-hand side cannot change its sign upon specifying $V(\phi)$, $\alpha$ and $c_\rho$. That is, the kinetic regime cannot be connected with the potential regime during the evolution of the system. As a consequence, the TDiff theory cannot lead to oscillations after a slow-roll phase, whenever the minimum potential energy vanishes, i.e. $V_{\rm min}=0$. This is rather significant compared to the Diff phenomenology, where the above equation is trivially satisfied. Notice that this conclusion has been reached only for this particular framework in which there is a universal volume factor for both the kinetic and potential terms in action \eqref{eq:actionphi}, a power-law for this volume function \eqref{eq:plaw} and $V_{\rm min}=0$.
Therefore, we shall carefully study which models are in principle compatible with slow-roll, i. e. potential domination. Moreover, the field should still satisfy the constraint after the end of inflation, new behaviors for the post-inflationary phase are expected.

In light of this novel phenomenology, in the present section we aim to comprehend better the involved dynamics during the inflationary epoch and immediately afterwards. That is, we want to address the study of the resulting TDiff dynamical systems. For this purpose, let us now analyze the ODEs system that rules the evolution of the involved fields, namely the inflaton $\phi$ and $Y$, and the universe via the Hubble parameter $H$. We recall that $\alpha V(\phi)>0$, in order to have a positive potential energy density in $\rho$ \eqref{eq:rho TDiff plaw} during slow-roll. For the sake of simplicity, we will consider a non-vanishing as well as positive potential, so that $\alpha>0$.

In this case, we have three equations, namely: the EoM \eqref{eq:EoM TDiff} on a FLRW background \eqref{eq:ds2 FRLW}, the Friedmann equation \eqref{eq:FLRW}, and the TDiff constraint \eqref{eq:TDiff const ps}. For later convenience, the last equation can be differentiated, so that the ODEs system reads
\begin{subequations}
\begin{empheq}[left=\empheqlbrace]{align}
   \ddot{\phi} + \left[3H+(1-2\alpha)H_Y\right]\dot\phi+V^{\prime}\left(\phi\right) & =0 , \label{eq:ODEeom} \\
    \frac{8\pi G}{3} Y^{1-2\alpha} \left[(1-\alpha) \dot{\phi}^2+2 \alpha V(\phi)\right] & =H^2 ,\label{eq:ODEfried} \\
    \left[\ddot{\phi}-V^\prime(\phi)\right]\dot{\phi}-2\alpha H_Y\left[\frac{1}{2} \dot{\phi}^2-V(\phi)\right] & = 0 ,\label{eq:ODEconst}
\end{empheq}
\end{subequations}

%v=-2*(1+2u)/1-2\alpha; A^2=12*u^2*(1-(1-\alpha*v)*u)/(\alpha*v-2*u); z=-1/3/u*(u*(u-1)+(1-2*\alpha)*v*u+1/6*A^2)
where we have defined the quantity
\begin{equation}
    H_Y=\frac{\dot{Y}}{Y},
    \label{eq:def HY}
\end{equation}
Some general remarks can be made about this system of coupled equations. On the one hand, the EoM \eqref{eq:ODEeom} still represents a damped oscillator but, unlike the general relativistic case, it now features two contributions to the friction term: one related to the expansion rate of the universe $H$ and the other related to the TDiff theory via $H_Y$. Therefore, the EoM presents two different limiting regimes: I) the case $H\gg H_Y$ yields the standard damped oscillator for $\phi$ in the Diff theory \cite{Kolb:1990vq,DiMarco:2024yzn}; and II) the opposite case $H_Y\gg H$ provides novel phenomenology. We will refer to the latter case as {\it strong TDiff regime} (STR), which is equivalent to neglecting the cosmological expansion $H$. As we will explicitly see below, we expect to reach the STR after inflation when $H$ start decreasing. On the other hand, we stress that there is also a sign change in the EoM \eqref{eq:ODEeom}, depending on whether $\alpha$ is below or above $1/2$. This sign change alters the behavior of the solution, specifically, the coefficient of the friction term $\dot\phi$.

The previous remarks are completely general. Nevertheless, throughout the rest of the section we will focus on the case of a quadratic potential $V\propto\phi^2$ as is usually done in reheating analysis since it is a good approximation for more general potential around their minimum. In particular, we will study in detail the STR, obtain the resulting phase portraits and perform a numerical analysis of a particular example.
\subsection{Strong TDiff regime}\label{sec:Strong TDiff regime}
Let us then consider a mass-term potential 
\begin{equation}
    V(\phi)=\frac{1}{2}m^2\phi^2,\label{eq:qua}
\end{equation}
and study the STR, that is $H_Y\gg H$. Focusing on the field content, in the first place, we consider the conservation equation \eqref{eq:cons} and neglect the cosmic expansion, which implies $\dot\rho=0$ in this regime; therefore, taking into account equation \eqref{eq:rho TDiff plaw}, this implies 
\begin{equation}
    \rho = Y^{1-2\alpha}\left[(1-\alpha)\dot{\phi}^2+\alpha\, m^2\phi^2\right]  = c_1, \label{eq:STDiff cons0}
\end{equation}
with $c_1$ a constant parameter. In the second place, the constraint equation \eqref{eq:TDiff const ps} for potential \eqref{eq:qua} is
\begin{equation}
    Y^{-2\alpha}(\dot{\phi}^2-m^2\phi^2)  = c_2,\label{eq:STDiff const0}
\end{equation}
where $c_2=-c_\rho/(1-2\alpha)$. Now, we can introduce dimensionless variables
\begin{equation}
    \hat{t}= mt,\quad \hat{\phi}= A\,{\phi},\quad \hat{Y}= B\,Y;
\end{equation}
with $m$, $A$ and $B$ being constant parameters, and normalize the equations \eqref{eq:STDiff cons0} and \eqref{eq:STDiff const0} by fixing the constants. Introducing the variables into \eqref{eq:STDiff cons0}, we find $B=\left(m^2A^{-2}{c_1}^{-1}\right)^{1/(1-2\alpha)}$ to normalize the right-hand-size; whereas for equation \eqref{eq:STDiff const0}, the right-hand side can be normalized if the parameters satisfy $|c_2|=(B^{\alpha}\,m/A)^2$, so we choose $A=m/(c_1^\alpha\,|c_2|^{-\alpha+1/2})$. Thus, we obtain the following algebraic system for the phase space $\{\phi,\dot{\phi}\}$ from equations:
\begin{subequations}
\begin{empheq}[left=\empheqlbrace]{align}
    Y^{1-2\alpha}\left[(1-\alpha)\dot{\phi}^2+\alpha\phi^2\right] & = 1 , \label{eq:STDiff cons} \\
    Y^{-2\alpha}(\dot{\phi}^2-\phi^2) & = c_2;\label{eq:STDiff const}
\end{empheq}
\end{subequations}
where now we have denoted $\dot{}=d/d\hat{t}$, removed the hats to ease the notation, and considered $c_2$ normalized to $\pm1,0$. Solving now \eqref{eq:STDiff cons} and \eqref{eq:STDiff const} for $\phi^2$ and $\dot\phi^2$, we find
\begin{subequations}
\begin{align}    
    \phi^2 & = Y^{2\alpha-1}\left[1+c_2(\alpha-1)Y\right] ,     \label{eq:cmap phi}\\
        \dot{\phi}^2 & = Y^{2\alpha-1}\left[1+c_2\alpha Y\right]   .  \label{eq:cmap dphi}
\end{align}
\end{subequations}
Noting that the left-hand-side of equations \eqref{eq:cmap phi} and \eqref{eq:cmap dphi} is non-negative, the expression between brackets is bounded from below; this gives us information about the allowed values for $Y$. Table \ref{tab:Bound g} represents a summary of these values:
\begin{table}[H]
    \centering
    \begin{tabular}{c|c|c|c}
    \hline
    \hline
     $Y_{\rm max}$ & $c_2=-1$ & $c_2=+1$ & $c_2=0$ \\
    \hline
     $\alpha>1$  	&	$\alpha^{-1}$	& $-$ &	$-$	\\
    \hline
     $\alpha<1$  	&	$\alpha^{-1}$	&	$(1-\alpha)^{-1}$	&	$-$	\\
    \hline
    \hline
    \end{tabular}
    \caption{Upper bound for the field $Y$, the lower bound is always $0$.}
    \label{tab:Bound g}
\end{table}
On the other hand, it is possible to find an expression for $\dot Y$ when differentiating $\phi$ in \eqref{eq:cmap phi} and substituting $\dot\phi$ \eqref{eq:cmap dphi}. If we define $\mathcal{F}(Y)=\phi^2(Y)$, these expressions yield
\begin{equation}
     \dot{Y} =\frac{2\phi\dot{\phi}}{\mathcal{F}^\prime(Y)}=\frac{2Y}{(\alpha-1)\,r_\phi^2+\alpha}r_\phi
    \label{eq:dotY},
\end{equation}
where we have
\begin{equation}
    \mathcal{F}^\prime(Y)= 2Y^{2\alpha-1}\left[\left(\alpha-\frac{1}{2}\right)Y^{-1}+c_2\alpha(\alpha-1)\right]
    \label{eq:F'}
\end{equation}
and where the final expression has been recast in terms of the ratio
\begin{equation}
    r_\phi=\frac{\dot{\phi}}{\phi}
    \label{eq:rphi}
\end{equation}
for later convenience.
The roots of the function $\mathcal{F}^\prime(Y)$ \eqref{eq:F'} are located at
\begin{equation}
     Y_\infty = -\frac{2\alpha-1}{2\alpha(\alpha-1)c_2}
     \label{eq:roots F(Y)}.
\end{equation}
For this value, the function $\dot Y$ diverges. This fact will have consequences in the phase portraits, as we shall see below.

Finally, the time evolution of $Y$ can be obtained by substituting equations \eqref{eq:cmap phi}, \eqref{eq:cmap dphi} and \eqref{eq:F'} into the derivative \eqref{eq:dotY}, so that
\begin{figure*}[t]
    \centering
    \includegraphics[width=\linewidth]{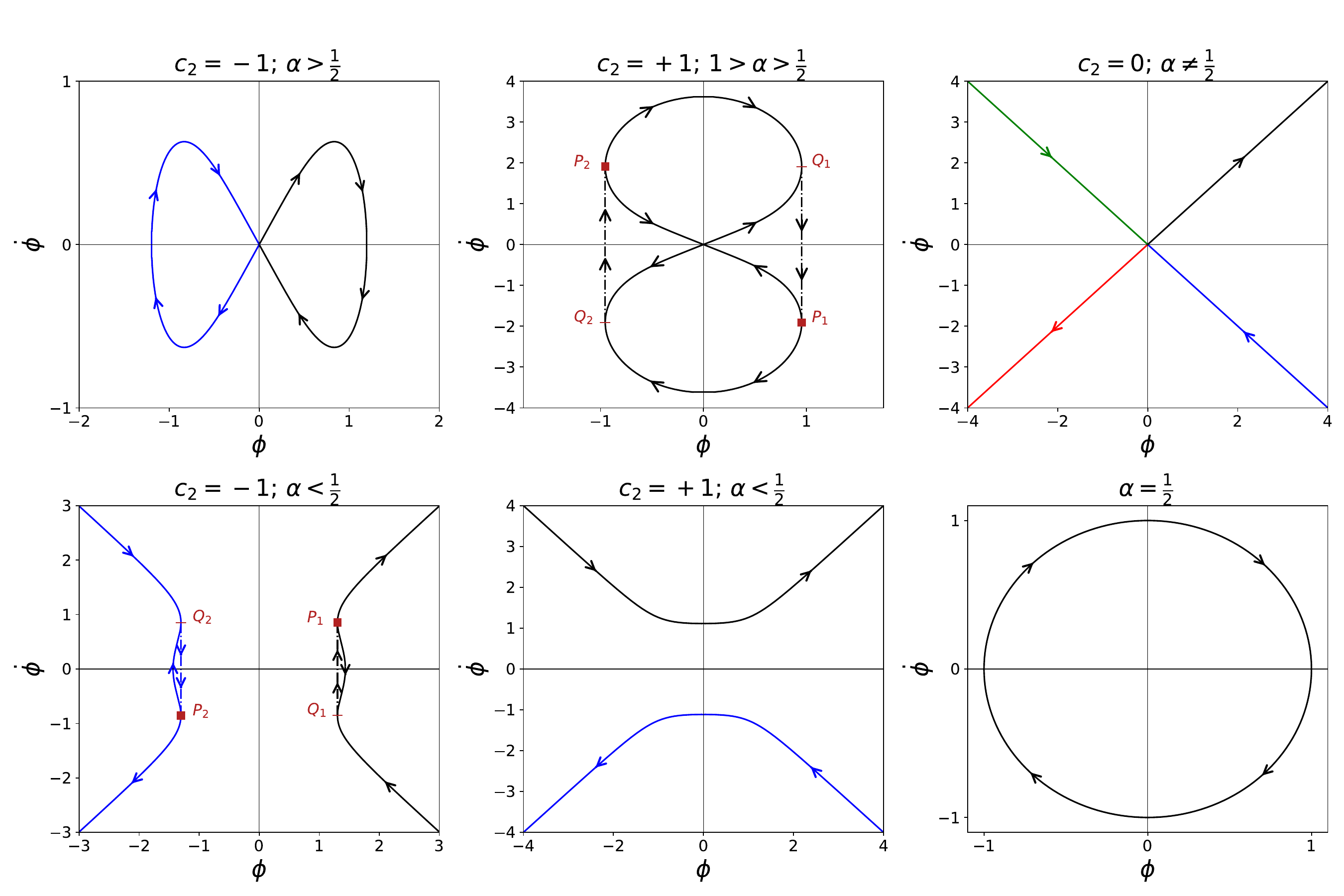}
    \caption{Phase portraits in the $\phi$--$\dot\phi$ plane, featuring $\{c_2,\alpha\}$ for each model. Left column: potential domination $c_2=-1$. Center column: kinetic domination $c_2=+1$. Right column: on top right, the limiting case $c_2=0$; on bottom right, the Diff oscillator case. The direction of the arrow indicates the evolution of the field regarding the sign of the velocity. The points marked as $P_i$ and $Q_i$ ($i=1,2$) represent bifurcation and ``brick-wall'' points, respectively. The different colors identify branches which cannot be connected.}
    \label{fig:Phase portraits}
\end{figure*}
\begin{equation}
    dt = \pm\frac{\left(\alpha-\frac{1}{2}\right)+c_2\alpha(\alpha-1)Y}{\sqrt{1+c_2(2\alpha-1)Y+c_2^2\alpha(\alpha-1)Y^2}}\frac{dY}{Y}
    \label{eq:dt cmap}.
\end{equation}
This expression can be integrated to determine whether certain points can be reached in a finite period of cosmological time, especially useful to understand the phase portraits.
\subsection{TDiff phase portraits in the STR}\label{sec:portraits}
We shall now examine the rich phenomenology of the STR ($H_Y\gg H$) in which we will neglect the universe expansion. To do so, we have plotted in figure \ref{fig:Phase portraits} the phase portraits of the solutions \eqref{eq:cmap phi} and \eqref{eq:cmap dphi}, namely the velocity $\dot\phi$ vs. the field $\phi$. As a brief reminder, we have also included in figure \ref{fig:Phase portraits} (bottom right panel) the phase portrait of the Diff harmonic oscillator  corresponding to  the EoM $\ddot\phi+\phi=0$ for the sake of comparison. As expected, the {\it position} $\phi$  moves from one end to the other, completely stopping and changing the direction of \textit{velocity} $\dot\phi$, without any kind of damping. In other words, the $\dot\phi$-intercepts represent the turning points.

Keeping this in mind, let us now discuss the TDiff phase portraits in terms of the classification $\{c_2,\alpha\}$. We still bear in mind the choice of $\alpha$ above or below $1/2$, as seen before, to study the TDiff results.
\subsubsection{Potential domination: $c_2=-1$}\label{sec:c2 -1}
Firstly, we address the relevant case for inflation. Starting now with $\alpha>1/2$ in figure \ref{fig:Phase portraits}, the top left panel, we can compare it to the Diff oscillator case, the bottom right panel. In contrast to a perfect circle, the TDiff damping makes the field tend to the origin. Thus, if the field starts with $\{\phi,\dot\phi\}>0$, it will follow the black line, gaining velocity, until it starts to slow down and stops at the value $Y_{\rm max}=\alpha^{-1}$ (see table \ref{tab:Bound g}). Once there, the field will change the direction of its movement, as the standard harmonic oscillator would do at the turning points.
After that, the field will continue to the origin $\phi=0$; however, the evaluation of the time \eqref{eq:dt cmap} for $Y\ll1$ (in order to have $\phi\to0$, according to equation \eqref{eq:cmap phi}), provides:
\begin{align}
   t\sim\ln Y\to\infty
    \label{eq:dt c2}.
\end{align}
Thus, the field requires an infinite period of time to reach the origin. We remark that we obtain an exponential asymptotic behavior for $Y$ if we invert the above relation, which will newly appear later.

The blue branch in the panel of figure \ref{fig:Phase portraits} just represents the reflected image of the above described motion, so both branches are disconnected from one another at the origin.

We now turn our attention to the case with $\alpha<1/2$ in figure \ref{fig:Phase portraits}, the bottom left panel. Let the field fall to smaller values, following the black branch, until it reaches the point $Q_1$, where it can no longer continue. This corresponds to the value of $Y_\infty$ \eqref{eq:roots F(Y)}. The only possible way out requires then the sign change of $\dot\phi$, represented by vertical dashed lines, while conserving the energy, as though it were an elastic collision. Thus, following \cite{Cline:2025tmd},  we will refer to this event as a {\it brick-wall} point.
Afterwards, the field reaches a bifurcation point $P_1$, where two trajectories are possible: (i) either $\phi$ moves away to infinite positive values, or (ii) it reaches some maximum value at $Y_{\rm max}=\alpha^{-1}$, it stops and then returns to the brick-wall point $Q_1$. The direction of the phase space flow allows the existence of a closed cycle, unless the bifurcation point $P_1$ leads the field to diverge toward infinity. Regarding the required time to reach infinity, the expression is also given by equation \eqref{eq:dt c2}, so it can never reach it within a finite time interval.
On the other hand, 
the blue branch represents again an inverted motion with the brick-wall point $Q_2$ and the bifurcation point $P_2$.

Note that brick-wall points do not introduce discontinuities in the cosmological observables as the energy density and pressure are quadratic in the velocity $\dot{\phi}$ in equations \eqref{eq:rho TDiff plaw} and \eqref{eq:p TDiff plaw}. Nevertheless, the bifurcation points compromise the predictability of the theory, as we cannot know in advance the chosen field trajectory at those points. 
\subsubsection{Kinetic domination: $c_2=+1$}\label{sec:c2 +1}
For the sake of completeness, we shall briefly review the rest of scenarios. Continuing now with the kinetic domination, we shall restrict to values $1>\alpha>1/2$, so that there is a maximum $Y_{\rm max}$ (see table \ref{tab:Bound g}). The behavior of the case $\alpha>1$ shall be studied in further research works. The analyzed case is pictured in figure \ref{fig:Phase portraits}, the top center panel. In this case,  if the field starts with maximum $\dot\phi>0$ at the origin, it will grow while loosing velocity, until it reaches the brick-wall point $Q_1$, corresponding to the value $Y_\infty$ \eqref{eq:roots F(Y)}. Now, $\dot\phi$ suffers the already mentioned sign change and the field reaches a bifurcation point $P_1$ with two possible trajectories: (i) either $\phi$ goes to the origin while slowing down, which implies that the field tends to zero ($Y\ll1$ due to equation \eqref{eq:cmap phi} and according to \eqref{eq:dt c2} the field never reaches the origin); or (ii) $\phi$ can cross the origin instead with $Y_{\rm max}=(1-\alpha)^{-1}$, as mentioned in table \ref{tab:Bound g}. For this last option, $\phi$ reaches another brick-wall point $Q_2$ and then it instantly changes the direction of movement.
After this, we likewise encounter another bifurcation point $P_2$: the system can either return to the origin or close the cycle. As we mentioned before, there is no predictability of this theory at those bifurcation points.

The other case, with $\alpha<1/2$, in figure \ref{fig:Phase portraits}, at the bottom center panel, can be more easily described. Here, for example, we have a field starting from negative values, represented by a black line, that slows down as it reaches the origin at $Y_{min}=(1-\alpha)^{-1}$. After crossing it, the field starts to gain velocity and finally moves away. The other blue branch is equivalent, but the direction of movement has been inverted. Additionally, the time expression is the one given by \eqref{eq:dt c2}, being in this case $\phi\to\infty$ for $Y\to0$, according to equation \eqref{eq:cmap phi} In other words, the field cannot reach infinity in a finite amount of time, either.
\subsubsection{Intermediate case: $c_2=0$}
Lastly, the phase portrait in figure \ref{fig:Phase portraits}, the top right panel, is valid for $\alpha$ above and below $1/2$, since $\phi=\dot\phi$. As before, if the field starts away from the origin, the initial conditions determine whether $\phi$ slowly tends to the origin or moves further away. In this case, we also have the time expression from equation \eqref{eq:dt c2}, so the field can never reach either the origin or infinity in a finite period of time (in equation \eqref{eq:cmap phi} we reach $\phi\to0$ for $\alpha>1/2$ when $Y\to0$ and for $\alpha<1/2$ when $Y\to\infty$; and we reach $\phi\to\infty$ for  $\alpha>1/2$ when $Y\to\infty$ and for $\alpha<1/2$ when $Y\to0$). Therefore, the branches cannot be connected and, consequently, they are represented with different colors.
\subsection{Example of TDiff dynamical system}\label{sec:Example of TDiff dynamical system}
The previous section was devoted to study the STR in different cases. However, we are now interested in analyzing the details of one specific example which could suit the inflationary phase in the context of slow-roll. We shall continue using the mass-term  potential \eqref{eq:qua}, as described in sections \ref{sec:Strong TDiff regime} and \ref{sec:portraits}. Substitution of this potential into equations \eqref{eq:eps plaw} and \eqref{eq:epseta plaw} yields
\begin{equation}
      \varepsilon  \simeq
      \frac{1}{16\pi G\alpha^2} \frac{Y^{2\alpha-1}}{\phi^2}
      \simeq
      \frac{1}{2\alpha}
        \eta  
    \label{eq:TDiff slr par},
\end{equation}
which successfully recovers the well-known Diff expression for $\alpha=1/2$, those are $\varepsilon\simeq\eta\simeq1/(4\pi G\phi^2)$. We still bear in mind that $\alpha V(\phi)>0$ (with both positive quantities), as before.

On the one hand, note that for $p=2$ we obtain $\alpha_p=1/2$ \eqref{eq:alpha pivot}. On the other hand, note also that the constraint \eqref{eq:STDiff const0} can be recast in this particular case in terms of $r_\phi$ \eqref{eq:rphi} as
\begin{equation}
     Y^{-2\alpha} \phi^2\left(r_\phi^2-1\right)=c_2,
     \label{eq:eps cota}
\end{equation}
where we have made the equations dimensionless again. According to our classification in section \ref{sec:Strong TDiff regime}, $c_2$ in equation \eqref{eq:eps cota} should be negative in order to obtain potential domination, so we can impose the negative sign on the expression in parentheses in equation \eqref{eq:eps cota}. Before doing so, we can apply the definition of $\varepsilon$ \eqref{eq:Diff epsilon} without focusing on the slow-roll regime; thus, substituting \eqref{eq:FLRW}, \eqref{eq:cons}, \eqref{eq:rho TDiff plaw} and \eqref{eq:p TDiff plaw} into equation \eqref{eq:Diff epsilon} it is possible to write
\begin{equation}
    \varepsilon = \frac{3}{2}\frac{r_\phi^2}{(1-\alpha)r_\phi^2+\alpha}\quad \Leftrightarrow \quad r_\phi^2=\frac{\alpha}{\frac{3}{2\varepsilon}+\alpha-1}
    \label{eq:eps rphi}.
\end{equation}
We can now substitute the expression on the right into the constraint \eqref{eq:eps cota} and impose $r_\phi^2-1<0$. This procedure yields the condition
\begin{equation}
    \varepsilon<\frac{3}{2},
    \label{eq:bound eps}
\end{equation}
for any $\alpha>0$. Thus, we have found an upper bound for the first slow-roll parameter.
In contrast to the last section, the cosmological expansion $H$ can now dominate against $H_Y$, so  the ODEs system \eqref{eq:ODEeom}--\eqref{eq:ODEconst} shall be numerically solved. However, as we have seen in Section \ref{sec:c2 -1}, the potential domination case could present some peculiar brick-wall and bifurcation points, if the STR is ever reached, which may eventually affect our solutions in the form of numerical singularities. For this reason, we have restricted ourselves to  the simplest case with $\{c_2=-1,\alpha>1/2\}$, which does not exhibit any type of discontinuity. We stress that for this parameter choice no instabilities in the scalar sector shall be expected during the whole evolution, as the effective speed of sound \eqref{eq:cs2} takes values in the $1>c_s^2>0$ range. In fact, initial conditions establish a potential dominance scenario, i.e. $V/X>1$, which is preserved in the
evolution by the constraint \eqref{eq:TDiff const ps}. We can see from Figure \ref{fig:cs2},  that the region with $\{\alpha>1/2,V/X>1\}$ indeed  corresponds to the mentioned values $1>c_s^2>0$.

Other choices of parameter space should be carefully analyzed in future research, but it is worth mentioning that superluminal propagation with $c_s^2>1$ does not necessarily generate causal paradoxes in this type of theories, as shown in reference \cite{Babichev:2007dw}.
\begin{figure*}[!t]
    \centering
    \includegraphics[width=\linewidth]{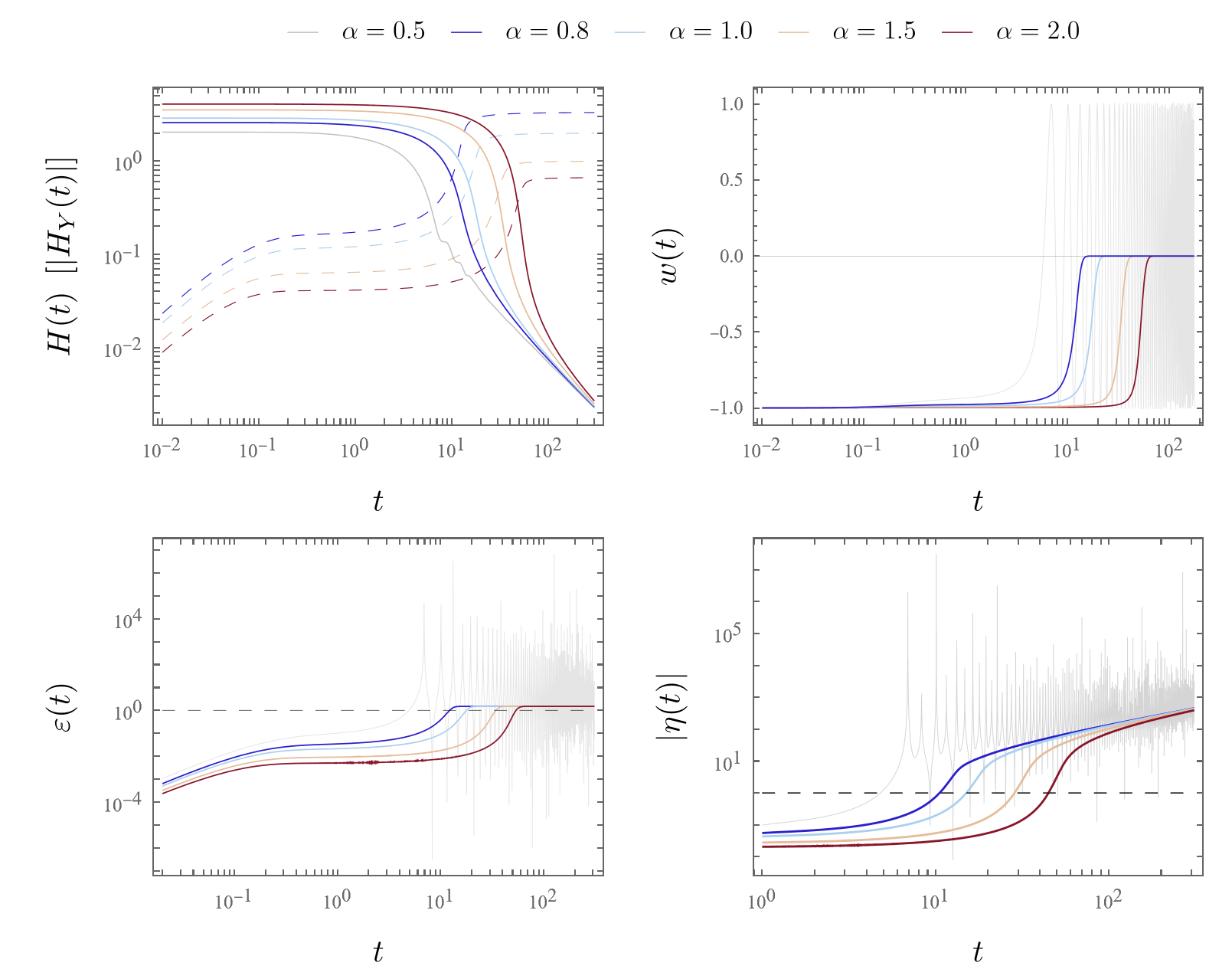}
    \caption{Numerical solutions of the dimensionless ODEs system \eqref{eq:ODEeom}--\eqref{eq:ODEconst}, showing the time evolution for representative values of $ \alpha$. Top left panel: Hubble parameter $H$ (solid line) and $|H_Y|=\left|\frac{\dot{Y}}{Y}\right|$ (dashed line). Top right panel: equation of state $w$. Bottom left panel: first slow-roll parameter $\varepsilon$ (the blurred brown oscillations seem to be numerical noise). Bottom right panel: second slow-roll parameter $|\eta|$. In these two last panels the dashed line indicates the condition of the end of slow-roll. In every panel the solid gray curve corresponds to the Diff case behavior $\left(\alpha=1/2\right)$, which always features oscillations.}
    \label{fig:panels}
\end{figure*}
\subsubsection{Numerical analysis}\label{sec:Numerical analysis}
The ODEs system  \eqref{eq:ODEeom}--\eqref{eq:ODEconst} has been numerically solved for different values of $\alpha>1/2$ and for the Diff case $\left(\alpha=1/2\right)$, with initial conditions
\begin{equation}
    a(0)= 1,\quad Y(0)=1,\quad \phi(0)=1,\quad \dot{\phi}(0)= 0.
\end{equation}
The results are shown in figure \ref{fig:panels}. Starting with the top left panel, the Hubble parameter $H$ and $|H_Y|$ feature a rather distinct behavior over time. On the one hand, in the Diff case, $H$ is nearly constant during slow-roll at early times, as expected during inflation. Then, it quickly decreases after the end of slow-roll with small amplitude oscillations corresponding to the oscillations around the potential minimum. The TDiff curves showcase a similar behavior, but in contrast to the previous case, they do not oscillate in the long term. This fact is related to the lack of oscillations due to the TDiff constraint \eqref{eq:STDiff const0}. Despite this, all curves seemingly converge towards the Diff case at late times independently of $\alpha$.

On the other hand, in all the cases considered, $|H_Y|$ is subdominant during the slow-roll phase, but then increases over time while $H$ decreases. At late times, $|H_Y|$ clearly becomes larger than $H$ and tends to a constant which depends on $\alpha$. Thus, in the post-inflationary phase the system enters  the strong TDiff regime described in  section \ref{sec:Strong TDiff regime}.

Moving to the top right panel in figure \ref{fig:panels}, we have computed the resulting inflaton EoS parameter \eqref{eq:w}. The Diff EoS parameter is just $w_\text{Diff}\simeq-1$ during slow-roll and it starts oscillating around zero after the end of inflation as it corresponds to the quadratic potential \cite{Turner:1983he}; whereas every TDiff EoS parameter shows the behavior $w_\text{TDiff}\to 0$ at late times. This is actually a quite remarkable result, since TDiff quadratic models apparently have a natural matter behavior for the inflaton $\phi$ at late times, for any $\alpha$ value. Note also that this very behavior would be equivalent to averaging Diff oscillations \cite{Turner:1983he}.

Finally, the last two panels contain information about the first and second slow-roll parameters. At early times, both $\varepsilon$ and $\eta$ are small, as expected during inflation, nevertheless, their long-term evolution is rather different. For instance, the bottom left panel in figure \ref{fig:panels} shows that $\varepsilon$ features a constant behavior for any $\alpha$ at late times, apparently going to the same constant value; whereas $\eta$ grows over time in the bottom right panel.

The above discussion suggests the search for an asymptotic behavior at late times that explains the post-inflationary epoch. We will now explore this idea under particular approximations.
\subsubsection{Asymptotic behavior: post-inflationary phase}
The assumption of reaching a strong TDiff regime, described in section \ref{sec:Strong TDiff regime}, is now confirmed. Indeed, during inflation the standard cosmological expansion $H$ dominates but, in light of the figure \ref{fig:panels}, $H$ becomes negligible in the long term compared to $H_Y$. Thus, the post-inflationary phase is then dominated by $H_Y$, consequently reaching the STR. Let us accordingly propose the following Ansatz at late times
\begin{equation}
    |H_Y|\sim\text{const.}\gg H,
    \label{eq:Ansatz}
\end{equation}
as seen on top left panel in figure \ref{fig:panels}.

The first immediate consequence is derived thanks to the definition of $H_Y$ \eqref{eq:def HY}, that is, $Y$ behaves asymptotically as
\begin{equation}
    Y\sim e^{H_Yt}
    \label{eq:Y asymp},
\end{equation}
as anticipated in section \ref{sec:c2 -1} via the equation \eqref{eq:dt c2}. In addition, by substituting the approximation \eqref{eq:Ansatz} into the EoM \eqref{eq:ODEeom} and making use of the quadratic potential, we obtain the following dimensionless, second-order linear constant-coefficient ODE,
\begin{equation}
    \ddot\phi +\left(1-2\alpha\right)H_Y\dot\phi+\phi = 0
    \label{eq:ODE asymp}.
\end{equation}
The general solution is a simple linear combination of exponential
\begin{equation}
\phi\sim c_1e^{\lambda_{+}t}+c_2e^{\lambda_{-}t}
    \label{eq:phi asym},
\end{equation}
where we have defined
\begin{equation}
  \lambda_{\pm} = \frac{1}{2}\left(-\theta\pm\sqrt{\theta^2-4}\right) ,
\label{eq:eigenval}
\end{equation}
and where
\begin{equation}
    \theta = \left(1-2\alpha\right)H_Y.
    \label{eq:theta}
\end{equation}

In order to determine the constant $H_Y$, let us consider the two possible cases $H_Y<0$ and $H_Y>0$. We may write the ratio $r_\phi^2$ \eqref{eq:rphi} by combining equations \eqref{eq:cmap phi} and \eqref{eq:cmap dphi}, so that
\begin{subequations} \label{eq:ratio}
\begin{empheq}[left={r_\phi^2 = \dfrac{1 + c_2 \alpha Y}{1 + c_2 (\alpha - 1) Y} \sim \empheqlbrace}]{align}
    1 &,\quad H_Y < 0 \label{eq:ratio_posi} \\
    \dfrac{\alpha}{\alpha - 1} &,\quad H_Y > 0 \label{eq:ratio_negati}
\end{empheq}
\end{subequations}
where we have used \eqref{eq:Y asymp} with the corresponding sign of $H_Y$ in the exponential. Furthermore, we can now substitute these asymptotic results \eqref{eq:ratio_posi} or \eqref{eq:ratio_negati} into \eqref{eq:dotY} and determine the $H_Y$ constant \eqref{eq:def HY}
\begin{subequations}\label{eq:HY}
\begin{empheq}[left={H_Y \sim \empheqlbrace}]{align}
    -\dfrac{2r_\phi}{1-2\alpha} = \dfrac{2}{1-2\alpha} &,\quad H_Y < 0 \label{eq:Hg positi}\\
    \dfrac{r_\phi}{\alpha} = +\dfrac{1}{\sqrt{\alpha(\alpha - 1)}} &,\quad H_Y > 0 \label{eq:Hg_negati}
\end{empheq}
\end{subequations}
where, in the last step, we have chosen $\text{sign}(r_\phi)$ to match the sign assumptions for $H_Y$.

Notwithstanding the above, the case which we are addressing in this approximation is the one shown on top left panel in figure \ref{fig:Phase portraits}, i. e. $\{c_2=-1,\alpha>1/2\}$. As the field rolls down to the origin, the plot implies that $r_\phi<0$. This fact makes unviable the solution with $H_Y>0$ \eqref{eq:Hg_negati}. For this reason, the only valid solution must be the negative one \eqref{eq:Hg positi}.

We are now able to explain the asymptotic behaviors of all the involved quantities in figure \ref{fig:panels}. For instance, substituting our solution \eqref{eq:Hg positi} into \eqref{eq:theta}, this result can be combined with \eqref{eq:eigenval} in order to find the time dependence of the scalar field \eqref{eq:phi asym}. The field then evolves as
\begin{equation}
    \phi\sim e^{-t}.
    \label{eq:phi asym fin}
\end{equation}

Recalling now the expression \eqref{eq:eps rphi} for the first slow-roll parameter, we can write 
\begin{equation}
   \varepsilon=\frac{3}{2}\frac{r_\phi^2}{(1-\alpha)r_\phi^2+\alpha}\sim\frac{3}{2},
   \label{eq:eps asymp}
\end{equation} 
where we have substituted the asymptotic value of $r_\phi^2$ \eqref{eq:ratio_posi}. The $\varepsilon$ parameter is indeed constant, no matter the value of $\alpha$, as anticipated in the bottom left panel in figure \ref{fig:panels}. In addition, it asymptotically goes to the same constant value. Moreover, it saturates the bound \eqref{eq:bound eps} after imposing the potential domination scenario, at the beginning of section \ref{sec:Example of TDiff dynamical system}.

Additionally, the substitution of the energy density \eqref{eq:rho TDiff plaw} and pressure \eqref{eq:p TDiff plaw} into the EoS parameter $w$ \eqref{eq:EoS phi} yields
\begin{equation}
     w= \frac{\alpha(r_\phi^2-1)}{(1-\alpha)r_\phi^2+\alpha}
     \label{eq:TDiff asymp EoS}.
\end{equation}
This parameter clearly vanishes at late times for any $\alpha>1/2$, given the asymptotic value \eqref{eq:ratio_posi}, that is, 
\begin{equation}
    w\sim0.
    \label{eq:w asymp}
\end{equation}
As a matter of fact, every model tends to a matter-dominated behavior, despite not being able to oscillate due to the constraint \eqref{eq:TDiff const ps}.

Furthermore, this matter behavior can be combined with the Friedmann equation \eqref{eq:FLRW}, so one readily finds
\begin{equation}
    H(t)\sim t^{-1}
    \label{eq:H asymp}.
\end{equation}
Thus, every TDiff model, independently of $\alpha$, yields the same time dependence. In the top left panel of figure \ref{fig:panels} we can appreciate this fact for the solid lines $H$.

Finally, the definition of $\eta$ \eqref{eq:Diff eta} yields in the asymptotic regime \eqref{eq:Ansatz}
\begin{equation}
    \eta=\varepsilon  -\frac{\ddot\phi}{\dot\phi H}\sim H^{-1}\sim t,
    \label{eq:eta asymp}
\end{equation}
This behavior completely agrees with the numerical results in the bottom right panel of figure \ref{fig:panels}, where we appreciate a linear growth at late times for $\eta$, independently of $\alpha$.

All of this information gives us a better understanding of the numerical solution for this model. We remark that the results truly separate two distinct limiting regimes: at early times, the standard friction term $H$ governs the EoM \eqref{eq:ODEeom}, but then $H_Y$ becomes dominant and rules the post-inflationary phase.
Notice that we are not addressing here the study of a plausible reheating phase. Such a study would entail a detailed analysis of the possible couplings of the inflaton field to the Diff fields.
\section{CONCLUSIONS}\label{sec:Conclusions}
In this work we have addressed the breaking of Diff invariance in the inflaton sector down to TDiff. We have explored the main changes caused by a TDiff coupling in the form of a single power-law function of the metric determinant. In particular, we have derived the relevant quantities during slow-roll  such as the slow-roll parameters and the number of $e$-folds, which differ from the Diff case.

We have also studied the primordial metric perturbations, taking into account that the symmetry breaking takes place only through the matter sector. This has allowed us to follow a standard quantization in order to compute the primordial power-spectrum. The obtained quantities such as the scalar amplitude and the spectral index showcase deviations with respect to the Diff case.

Furthermore, upon specifying a power-law form for the inflaton potential, we have derived the observable quantities and compared the predictions of the model with the ACT and  \textit{Planck} data sets. For potentials with exponents smaller than $p=2$, the TDiff coupling can improve the agreement with observations by reducing the tensor-to- scalar ratio and, in some cases, alleviate the tension below the $1\sigma$ limit. However, for powers above $p=2$, the improvement is only marginal.

On the other hand, we have also studied the post-inflationary phase in the TDiff context. In this regime the constraint equation becomes crucial for understanding the evolution of the inflaton. In particular, we have shown that, in our specific configuration, it prevents oscillations of the field after inflation. We have then analyzed in depth the subsequent dynamics for the quadratic potential, thus introducing the so-called strong TDiff regime that may be achieved in a post-inflationary phase. This regime could bring along the appearance of non-trivial events, such as brick-wall and bifurcation points. However, it should be emphasized that physical viability of the post-inflationary stage should  require not only the existence of a graceful exit from inflation, but also  a
viable mechanism that transfers the inflaton energy to radiation or to other sectors, and
the recovery of the standard hot Big Bang cosmology after the end of inflation. In the present work we have shown that a graceful exit is possible in this framework, but we leave the study of preheating and/or reheating for future works.

Moreover, we have conducted a detailed numerical analysis for the ODEs system with a quadratic potential in one specific case, namely the one compatible with slow-roll and with $\alpha>1/2$. The TDiff constraint prevents indeed the oscillations of the field and we obtain novel phenomenology. In light of these results, we have observed that at early times the cosmological expansion rate $H$ dominates over the TDiff contribution $H_Y$ in the friction term. As times goes, however, $H$ decreases and $H_Y$ grows; so, the strong TDiff regime is eventually achieved. We have investigated in detail the asymptotic behavior of the model. The results show that, independently of the TDiff parameter, inflation still has a graceful exit. The post-inflationary phase is certainly dominated by TDiff features. In addition, we have also observed a matter behavior at late times for the inflaton field regardless the value of $\alpha>1/2$.

Notwithstanding the above, further research can be performed witin the  TDiff framework of inflation. In particular, the rest of cases within the strong TDiff regime should be deeply discussed, especially the remaining case compatible with inflation with $\alpha<1/2$, which features brick-wall points as well as bifurcation points. Additionally, more complicated coupling functions for the volume element could be considered in the discussion together with potential forms beyond the simple power laws. In addition, different TDiff functions for the kinetic and potential parts would be an appealing possibility within this framework.
\section*{Acknowledgments}
We thank the anonymous referee for useful comments and suggestions. This work has been supported by the MICIN (Spain)
Project No. PID2022-138263NB-I00 funded by MICIU/AEI/10.13039/501100011033 and by ERDF/EU.
\appendix
\section{Propagating degrees of freedom and quadratic action}{\label{sec:ApA}}
In order to identify the number of propagating degrees of freedom present in the theory, let us rewrite the inflaton action
\begin{align}
 S_{\phi}^{\text{cov}} = \int d^4x \sqrt{g}H_K(Y)\left[X-V(\phi)\right]
    \label{actionap},
\end{align}
as follows
\begin{align}
 S_{\phi}^{\text{cov}} = \int d^4x \sqrt{g}\left\{H_K(\psi)\left[X-V(\phi)\right]+\lambda_0(\nabla_\mu A^\mu -\psi)\right\},
 \end{align}
where we have introduced the new scalar field $\psi$ through the Lagrange multiplier $\lambda_0$. The equations of motion for $A^\mu$  
imply $\lambda_0=$ const, whereas the equation of motion of the $\psi$ field implies
\begin{align}
 H_K'(\psi)[X-V(\phi)]=\lambda_0,
 \label{constpsi}
 \end{align}
which is nothing but the constraint equation \eqref{const}, where the Lagrange multiplier takes the constant value $\lambda_0=-c_\rho/2$. Since $\lambda_0$ is a 
constant then the $\nabla_\mu A^\mu$ term in the action is just a surface term so that we can write
\begin{align}
 S_{\phi}^{\text{cov}} = \int d^4x \sqrt{g}\left\{H_K(\psi)\left[X-V(\phi)\right]-\lambda_0 \psi\right\}.
 \end{align}
For instance,  in the power-law case $H_K(\psi)=\psi^{1-2\alpha}$, the constraint equation \eqref{constpsi} can be explicitly solved to give
\begin{align}
    \psi=\left[\frac{\lambda_0}{(1-2\alpha)(X-V)}\right]^{-\frac{1}{2\alpha}}.
\end{align}
Thus we see that the action  contains one local propagating degree of freedom $\phi$ and a global one   $\lambda_0$ which is fixed by the initial conditions. Notice that $\psi$ is just an auxiliary field. The presence of 
the global degree of freedom is a typical feature of theories with transverse diffeomorphisms. Thus, for instance, in unimodular gravity , the global degree of freedom appears as canonically conjugate
of the cosmological constant \cite{Henneaux:1989zc}, (see also \cite{Blas:2011ac} and  \cite{deCruzPerez:2025ytd} for extended TDiff models). 
 
Introducing  the auxiliary field back into the action, we find the final form in terms of the only propagating degree of freedom and the global one.
\begin{align}
 S_{\phi}^{\text{cov}} = \int d^4x \sqrt{g}\, P(X,\phi,\lambda_0),
\end{align} 
where, in the power-law case, we have
\begin{align}
 P(X,\phi,\lambda_0)=  2\alpha\frac{\lambda_0}{1-2\alpha}\left[\frac{1-2\alpha}{\lambda_0}(X-V)\right]^{\frac{1}{2\alpha}}.
\end{align} 
Thus we find that the action \eqref{actionap} actually corresponds to a family of k-inflation theories
parametrized by the global degree of freedom $\lambda_0$. This equivalence has been shown to work for large classes of TDiff theories
in \cite{BeltranJimenez:2025pho}. 
 
The corresponding total scalar quadratic action built out of the Einstein-Hilbert and the inflaton actions has been obtained for general  k-inflation models in \cite{Babichev:2007dw}, \cite{Langlois:2010xc} and reads
\begin{align}
 S_{R+\phi}^{(2)} = \frac{1}{2}\int d\eta\, d^3x\, \left[v'^2+c_s^2 v\nabla^2 v+\frac{z''}{z}v^2\right] \label{quadP},
\end{align} 
with
\begin{align}
    v=\frac{a\sqrt{P_X}}{c_s}\left(\delta\phi +\frac{\phi_0'}{\cal H}\Phi\right)
\end{align}
and
\begin{align}
    z=\frac{a\phi_0'\sqrt{P_X}}{c_s {\cal H}}.
\end{align}
Thus, in the power-law case, we find
\begin{align}
    P_X = \left.\frac{\partial P}{\partial X}\right\rvert_ {X_0,V_0}=\left[\frac{1-2\alpha}{\lambda_0}\left(X_0-V_0\right)\right]^{\frac{1}{2\alpha}-1}
\end{align}
and 
\begin{align}
    c_s^2=\frac{P_X}{P_X+2XP_{XX}},
\end{align}
which agrees with \eqref{eq:cs2}. 
The general quadratic action \eqref{quadP}, shows that the model only propagates one local scalar degree of freedom. As a matter of fact, there are no ghosts instabilities provided $P_X>0$ which is nothing but the condition $H_K(Y)>0$ which ensures the positivity of the kinetic term in 
the original action \eqref{actionap}.
On the other hand, the absence of gradient instabilities is also guaranteed
in models with $c_s^2>0$. Furthermore, notice that superluminal propagation with  $c_s^2>1$  does not necessarily generate  causal paradoxes in this type of theories 
as shown in \cite{Babichev:2007dw}.
%------------------------------------------------------------------------------------------
\bibliographystyle{elsarticle-num} 
\bibliography{biblio}

@article{Riotto:2002yw,
    author = "Riotto, Antonio",
    editor = "Dvali, G. and Perez-Lorenzana, Abdel and Senjanovic, G. and Thompson, G. and Vissani, F.",
    title = "{Inflation and the theory of cosmological perturbations}",
    eprint = "hep-ph/0210162",
    archivePrefix = "arXiv",
    reportNumber = "DFPD-TH-02-22",
    journal = "ICTP Lect. Notes Ser.",
    volume = "14",
    pages = "317--413",
    year = "2003"
}

@article{Langlois:2010xc,
    author = "Langlois, D.",
    title = "{Lectures on inflation and cosmological perturbations}",
    eprint = "1001.5259",
    archivePrefix = "arXiv",
    primaryClass = "astro-ph.CO",
    doi = "10.1007/978-3-642-10598-2_1",
    journal = "Lect. Notes Phys.",
    volume = "800",
    pages = "1--57",
    year = "2010"
}

@article{Maroto:2023toq,
    author = "Maroto, Antonio L.",
    title = "{TDiff invariant field theories for cosmology}",
    eprint = "2301.05713",
    archivePrefix = "arXiv",
    primaryClass = "gr-qc",
    reportNumber = "IPARCOS-UCM-23-006",
    doi = "10.1088/1475-7516/2024/04/037",
    journal = "JCAP",
    volume = "04",
    pages = "037",
    year = "2024"
}

@article{Jaramillo-Garrido:2024tdv,
    author = "Jaramillo-Garrido, Dar\'\i{}o and Maroto, Antonio L. and Mart\'\i{}n-Moruno, Prado",
    title = "{Symmetry restoration in transverse diffeomorphism invariant scalar field theories}",
    eprint = "2402.17422",
    archivePrefix = "arXiv",
    primaryClass = "gr-qc",
    reportNumber = "IPARCOS-UCM-24-013",
    doi = "10.1103/PhysRevD.110.044009",
    journal = "Phys. Rev. D",
    volume = "110",
    number = "4",
    pages = "044009",
    year = "2024"
}

@article{Cline:2025tmd,
    author = "Cline, James M.",
    title = "{Comment on ''Dark Energy from Time Crystals''}",
    eprint = "2502.19448",
    archivePrefix = "arXiv",
    journal = "",
    primaryClass = "hep-ph",
    month = "2",
    year = "2025"
}

@book{Kolb:1990vq,
    author = "Kolb, Edward W. and Turner, Michael S.",
    title = "{The Early Universe}",
    reportNumber = "FERMILAB-BOOK-1990-01",
    doi = "10.1201/9780429492860",
    isbn = "978-0-429-49286-0, 978-0-201-62674-2",
    publisher = "Taylor and Francis",
    volume = "69",
    month = "5",
    year = "2019"
}

@article{Alonso-Lopez:2023hkx,
    author = "Alonso-L\'opez, David and de Cruz P\'erez, Javier and Maroto, Antonio L.",
    title = "{Unified transverse diffeomorphism invariant field theory for the dark sector}",
    eprint = "2311.16836",
    archivePrefix = "arXiv",
    primaryClass = "astro-ph.CO",
    reportNumber = "IPARCOS-UCM-23-130",
    doi = "10.1103/PhysRevD.109.023537",
    journal = "Phys. Rev. D",
    volume = "109",
    number = "2",
    pages = "023537",
    year = "2024"
}

@article{Jaramillo-Garrido:2023cor,
    author = "Jaramillo-Garrido, Dar\'\i{}o and Maroto, Antonio L. and Mart\'\i{}n-Moruno, Prado",
    title = "{TDiff in the dark: gravity with a scalar field invariant under transverse diffeomorphisms}",
    eprint = "2307.14861",
    archivePrefix = "arXiv",
    primaryClass = "gr-qc",
    reportNumber = "IPARCOS-UCM-23-064",
    doi = "10.1007/JHEP03(2024)084",
    journal = "JHEP",
    volume = "03",
    pages = "084",
    year = "2024"
}

@article{Tessainer:2024ewm,
    author = "Tessainer, Diego and Maroto, Antonio L. and Mart\'\i{}n-Moruno, Prado",
    title = "{Multi-field TDiff theories for cosmology}",
    eprint = "2409.11991",
    archivePrefix = "arXiv",
    primaryClass = "gr-qc",
    reportNumber = "IPARCOS-UCM-24-042",
    doi = "10.1016/j.dark.2024.101769",
    journal = "Phys. Dark Univ.",
    volume = "47",
    pages = "101769",
    year = "2025"
}

@article{Planck:2018jri,
    author = "Akrami, Y. and others",
    collaboration = "Planck",
    title = "{Planck 2018 results. X. Constraints on inflation}",
    eprint = "1807.06211",
    archivePrefix = "arXiv",
    primaryClass = "astro-ph.CO",
    doi = "10.1051/0004-6361/201833887",
    journal = "Astron. Astrophys.",
    volume = "641",
    pages = "A10",
    year = "2020"
}

@article{Einstein:1919gv,
    author = "Einstein, Albert",
    title = "{Spielen Gravitationsfelder im Aufbau der materiellen Elementarteilchen eine wesentliche Rolle?}",
    journal = "Sitzungsber. Preuss. Akad. Wiss. Berlin (Math. Phys. )",
    volume = "1919",
    pages = "349--356",
    year = "1919"
}

@book{Mukhanov:2005sc,
    author = "Mukhanov, V.",
    title = "{Physical Foundations of Cosmology}",
    doi = "10.1017/CBO9780511790553",
    isbn = "978-0-521-56398-7",
    publisher = "Cambridge University Press",
    address = "Oxford",
    year = "2005"
}

@article{ACT:2025tim,
    author = "Calabrese, Erminia and others",
    collaboration = "ACT",
    title = "{The Atacama Cosmology Telescope: DR6 Constraints on Extended Cosmological Models}",
    eprint = "2503.14454",
    journal = "",
    archivePrefix = "arXiv",
    primaryClass = "astro-ph.CO",
    reportNumber = "FERMILAB-PUB-25-0157-PPD",
    month = "3",
    year = "2025"
}

@article{DiMarco:2024yzn,
    author = "Di Marco, Alessandro Di and Orazi, Emanuele and Pradisi, Gianfranco",
    title = "{Introduction to the Number of e-Folds in Slow-Roll Inflation}",
    eprint = "2408.01854",
    archivePrefix = "arXiv",
    primaryClass = "astro-ph.CO",
    doi = "10.3390/universe10070284",
    journal = "Universe",
    volume = "10",
    number = "7",
    pages = "284",
    year = "2024"
}

@article{DESI:2025zgx,
    author = "Abdul Karim, M. and others",
    collaboration = "DESI",
    title = "{DESI DR2 results. II. Measurements of baryon acoustic oscillations and cosmological constraints}",
    eprint = "2503.14738",
    archivePrefix = "arXiv",
    primaryClass = "astro-ph.CO",
    reportNumber = "FERMILAB-PUB-25-0169-PPD",
    doi = "10.1103/tr6y-kpc6",
    journal = "Phys. Rev. D",
    volume = "112",
    number = "8",
    pages = "083515",
    year = "2025"
}

@misc{Demirel2025PlanckConstraints,
  author       = {K. Demirel},
  title        = {Planck-constraints-r-vs-ns-plot},
  year         = {2025},
  howpublished = {\url{https://github.com/kdemirel/Planck-constraints-r-vs-ns-plot}},
}

@article{Martin:2013tda,
    author = "Martin, Jerome and Ringeval, Christophe and Vennin, Vincent",
    title = "{Encyclop{\ae}dia Inflationaris}: {Opiparous Edition}",
    eprint = "1303.3787",
    archivePrefix = "arXiv",
    primaryClass = "astro-ph.CO",
    doi = "10.1016/j.dark.2024.101653",
    journal = "Phys. Dark Univ.",
    volume = "5-6",
    pages = "75--235",
    year = "2014"
}

@article{Kallosh:2025rni,
    author = "Kallosh, Renata and Linde, Andrei and Roest, Diederik",
    title = "{Atacama Cosmology Telescope, South Pole Telescope, and Chaotic Inflation}",
    eprint = "2503.21030",
    archivePrefix = "arXiv",
    primaryClass = "hep-th",
    doi = "10.1103/d6gn-78hn",
    journal = "Phys. Rev. Lett.",
    volume = "135",
    number = "16",
    pages = "161001",
    year = "2025"
}

@article{SPT-3G:2025bzu,
    author = "Camphuis, E. and others",
    collaboration = "SPT-3G",
    title = "{SPT-3G D1: CMB temperature and polarization power spectra and cosmology from 2019 and 2020 observations of the SPT-3G Main field}",
    eprint = "2506.20707",
    journal = "",
    archivePrefix = "arXiv",
    primaryClass = "astro-ph.CO",
    reportNumber = "FERMILAB-PUB-25-0144-PPD",
    month = "6",
    year = "2025"
}

@article{Gao:2025onc,
    author = "Gao, Qing and Gong, Yungui and Yi, Zhu and Zhang, Fengge",
    title = "{Nonminimal coupling in light of ACT data}",
    eprint = "2504.15218",
    archivePrefix = "arXiv",
    primaryClass = "astro-ph.CO",
    doi = "10.1016/j.dark.2025.102106",
    journal = "Phys. Dark Univ.",
    volume = "50",
    pages = "102106",
    year = "2025"
}

@article{deCruzPerez:2025ytd,
    author = "de Cruz P{\'e}rez, Javier and Maroto, Antonio L.",
    title = "{{\ensuremath{\Lambda}}CDM from broken diffeomorphisms}",
    eprint = "2504.02541",
    archivePrefix = "arXiv",
    primaryClass = "gr-qc",
    reportNumber = "IPARCOS-UCM-25-022",
    doi = "10.1103/rxnw-gvyd",
    journal = "Phys. Rev. D",
    volume = "111",
    number = "12",
    pages = "123555",
    year = "2025"
}

@article{BeltranJimenez:2025puw,
    author = "Beltr{\'a}n Jim{\'e}nez, Jose and Borislavov Vasilev, Teodor and Jaramillo-Garrido, Dar{\'\i}o and Maroto, Antonio L. and Mart{\'\i}n-Moruno, Prado",
    title = "{K-nonizing}",
    eprint = "2509.07715",
    archivePrefix = "arXiv",
    journal = "",
    primaryClass = "gr-qc",
    reportNumber = "IPARCOS-UCM-25-044",
    month = "9",
    year = "2025"
}

@article{Maroto:2025ife,
    author = "Maroto, Antonio L. and Mart{\'\i}n-Moruno, Prado and Tessainer, Diego",
    title = "{Multi-field TDiff theories: the mixed regime case}",
    eprint = "2507.16616",
    archivePrefix = "arXiv",
    primaryClass = "gr-qc",
    journal = "",
    reportNumber = "IPARCOS-UCM-25-043",
    month = "7",
    year = "2025"
}

@article{Linde:1983gd,
    author = "Linde, Andrei D.",
    title = "{Chaotic Inflation}",
    doi = "10.1016/0370-2693(83)90837-7",
    journal = "Phys. Lett. B",
    volume = "129",
    pages = "177--181",
    year = "1983"
}

@article{Bezrukov:2007ep,
    author = "Bezrukov, Fedor L. and Shaposhnikov, Mikhail",
    title = "{The Standard Model Higgs boson as the inflaton}",
    eprint = "0710.3755",
    archivePrefix = "arXiv",
    primaryClass = "hep-th",
    doi = "10.1016/j.physletb.2007.11.072",
    journal = "Phys. Lett. B",
    volume = "659",
    pages = "703--706",
    year = "2008"
}

@article{Starobinsky:1980te,
    author = "Starobinsky, Alexei A.",
    editor = "Khalatnikov, I. M. and Mineev, V. P.",
    title = "{A New Type of Isotropic Cosmological Models Without Singularity}",
    doi = "10.1016/0370-2693(80)90670-X",
    journal = "Phys. Lett. B",
    volume = "91",
    pages = "99--102",
    year = "1980"
}

@article{Guth:1980zm,
    author = "Guth, Alan H.",
    editor = "Fang, Li-Zhi and Ruffini, R.",
    title = "{The Inflationary Universe: A Possible Solution to the Horizon and Flatness Problems}",
    reportNumber = "SLAC-PUB-2576",
    doi = "10.1103/PhysRevD.23.347",
    journal = "Phys. Rev. D",
    volume = "23",
    pages = "347--356",
    year = "1981"
}

@article{Unruh.40.1048,
  title = {Unimodular theory of canonical quantum gravity},
  author = {Unruh, W. G.},
  journal = {Phys. Rev. D},
  volume = {40},
  issue = {4},
  pages = {1048--1052},
  numpages = {0},
  year = {1989},
  month = {Aug},
  publisher = {American Physical Society},
  doi = {10.1103/PhysRevD.40.1048},
  url = {https://link.aps.org/doi/10.1103/PhysRevD.40.1048}
}

@article{Carballo-Rubio:2022ofy,
    author = "Carballo-Rubio, Ra\'ul and Garay, Luis J. and Garc\'\i{}a-Moreno, Gerardo",
    title = "{Unimodular gravity vs general relativity: a status report}",
    eprint = "2207.08499",
    archivePrefix = "arXiv",
    primaryClass = "gr-qc",
    doi = "10.1088/1361-6382/aca386",
    journal = "Class. Quant. Grav.",
    volume = "39",
    number = "24",
    pages = "243001",
    year = "2022"
}

@article{articleEllis,
author = {Ellis, George and van Elst, Henk and Murugan, Jeff and Uzan, Jean-Philippe},
year = {2011},
month = {10},
pages = {225007},
title = {On the trace-free Einstein equations as a viable alternative to General Relativity},
volume = {28},
journal = {Classical and Quantum Gravity},
doi = {10.1088/0264-9381/28/22/225007}
}

@article{Alvarez:2006uu,
    author = "Alvarez, E. and Blas, D. and Garriga, J. and Verdaguer, E.",
    title = "{Transverse Fierz-Pauli symmetry}",
    eprint = "hep-th/0606019",
    archivePrefix = "arXiv",
    reportNumber = "IFT-UAM-CSIC-05-43",
    doi = "10.1016/j.nuclphysb.2006.08.003",
    journal = "Nucl. Phys. B",
    volume = "756",
    pages = "148--170",
    year = "2006"
}

@article{Pirogov:2011iq,
    author = "Pirogov, Yu. F.",
    title = "{Unimodular bimode gravity and the coherent scalar-graviton field as galaxy dark matter}",
    eprint = "1111.1437",
    archivePrefix = "arXiv",
    primaryClass = "gr-qc",
    doi = "10.1140/epjc/s10052-012-2017-y",
    journal = "Eur. Phys. J. C",
    volume = "72",
    pages = "2017",
    year = "2012"
}

@article{Bello-Morales:2024vqk,
    author = "Bello-Morales, Antonio G. and Beltr\'an Jim\'enez, Jose and Jim\'enez Cano, Alejandro and Maroto, Antonio L. and Koivisto, Tomi S.",
    title = "{A class of ghost-free theories in symmetric teleparallel geometry}",
    eprint = "2406.19355",
    archivePrefix = "arXiv",
    journal = "", 
    primaryClass = "gr-qc",
    month = "6",
    year = "2024"
}

@article{Bello-Morales:2023btf,
    author = "Bello-Morales, Antonio G. and Maroto, Antonio L.",
    title = "{Cosmology in gravity models with broken diffeomorphisms}",
    eprint = "2308.00635",
    archivePrefix = "arXiv",
    primaryClass = "gr-qc",
    reportNumber = "IPARCOS-UCM-23-066",
    doi = "10.1103/PhysRevD.109.043506",
    journal = "Phys. Rev. D",
    volume = "109",
    number = "4",
    pages = "043506",
    year = "2024"
}

@article{Maroto:2024mkx,
    author = "Maroto, Antonio L. and Miravet, Alfredo D.",
    title = "{Transverse-diffeomorphism invariant gauge fields in cosmology}",
    eprint = "2402.18368",
    archivePrefix = "arXiv",
    primaryClass = "gr-qc",
    reportNumber = "IPARCOS-UCM-24-014",
    doi = "10.1103/PhysRevD.109.103504",
    journal = "Phys. Rev. D",
    volume = "109",
    number = "10",
    pages = "103504",
    year = "2024"
}

@article{Maroto:2024roe,
    author = "Maroto, Antonio L. and Miravet, Alfredo D.",
    title = "{Cosmic magnetic fields invariant under transverse diffeomorphisms}",
    eprint = "2407.04647",
    archivePrefix = "arXiv",
    primaryClass = "astro-ph.CO",
    doi = "10.1103/PhysRevD.110.063530",
    journal = "Phys. Rev. D",
    volume = "110",
    number = "6",
    pages = "063530",
    year = "2024"
}

@book{Baumann:2022mni,
    author = "Baumann, Daniel",
    title = "{Cosmology}",
    doi = "10.1017/9781108937092",
    isbn = "978-1-108-93709-2, 978-1-108-83807-8",
    publisher = "Cambridge University Press",
    month = "7",
    year = "2022"
}

@article{Liddle:1994dx,
    author = "Liddle, Andrew R. and Parsons, Paul and Barrow, John D.",
    title = "{Formalizing the slow roll approximation in inflation}",
    eprint = "astro-ph/9408015",
    archivePrefix = "arXiv",
    reportNumber = "SUSSEX-AST-94-8-1",
    doi = "10.1103/PhysRevD.50.7222",
    journal = "Phys. Rev. D",
    volume = "50",
    pages = "7222--7232",
    year = "1994"
}

@article{Turner:1983he,
    author = "Turner, Michael S.",
    title = "{Coherent Scalar Field Oscillations in an Expanding Universe}",
    reportNumber = "EFI-83-29-CHICAGO",
    doi = "10.1103/PhysRevD.28.1243",
    journal = "Phys. Rev. D",
    volume = "28",
    pages = "1243",
    year = "1983"
}

@article{Kim:2025dyi,
    author = "Kim, Jinsu and Wang, Xinpeng and Zhang, Ying-li and Ren, Zhongzhou",
    title = "{Enhancement of primordial curvature perturbations in R $^{3}$-corrected Starobinsky-Higgs inflation}",
    eprint = "2504.12035",
    archivePrefix = "arXiv",
    primaryClass = "astro-ph.CO",
    doi = "10.1088/1475-7516/2025/09/011",
    journal = "JCAP",
    volume = "09",
    pages = "011",
    year = "2025"
}

@article{Babichev:2007dw,
    author = "Babichev, Eugeny and Mukhanov, Viatcheslav and Vikman, Alexander",
    title = "{k-Essence, superluminal propagation, causality and emergent geometry}",
    eprint = "0708.0561",
    archivePrefix = "arXiv",
    primaryClass = "hep-th",
    reportNumber = "LMU-ASC-54-07",
    doi = "10.1088/1126-6708/2008/02/101",
    journal = "JHEP",
    volume = "02",
    pages = "101",
    year = "2008"
}

@article{Blas:2011ac,
    author = "Blas, Diego and Shaposhnikov, Mikhail and Zenhausern, Daniel",
    title = "{Scale-invariant alternatives to general relativity}",
    eprint = "1104.1392",
    archivePrefix = "arXiv",
    primaryClass = "hep-th",
    doi = "10.1103/PhysRevD.84.044001",
    journal = "Phys. Rev. D",
    volume = "84",
    pages = "044001",
    year = "2011"
}

@article{Henneaux:1989zc,
    author = "Henneaux, M. and Teitelboim, C.",
    title = "{The Cosmological Constant and General Covariance}",
    doi = "10.1016/0370-2693(89)91251-3",
    journal = "Phys. Lett. B",
    volume = "222",
    pages = "195--199",
    year = "1989"
}

@article{BeltranJimenez:2025pho,
    author = "Beltr{\'a}n Jim{\'e}nez, Jose and Borislavov Vasilev, Teodor and Jaramillo-Garrido, Dar{\'\i}o and Maroto, Antonio L. and Mart{\'\i}n-Moruno, Prado",
    title = "{K-nonizing}",
    eprint = "2509.07715",
    archivePrefix = "arXiv",
    primaryClass = "gr-qc",
    reportNumber = "IPARCOS-UCM-25-044",
    doi = "10.1016/j.physletb.2025.140118",
    journal = "Phys. Lett. B",
    volume = "872",
    pages = "140118",
    year = "2026"
}
%------------------------------------------------------------------------------------------
\end{document}